\UseRawInputEncoding
\documentclass[showpacs,twocolumn,superscriptaddress,nofootinbib]{revtex4-1} 

\usepackage{hyperref}
\hypersetup{colorlinks=true,linkcolor=blue,citecolor=blue}
\usepackage{tikz}
\usepackage{graphicx}
\usepackage{xcolor, soul}
\sethlcolor{green}
\usepackage{dcolumn}
\usepackage{bm}
\usepackage{color}
\usepackage{enumitem}
\usepackage{mathrsfs}
\usepackage{amsmath}
\usepackage{amssymb}
\usepackage{cleveref}
\usepackage{orcidlink}
\usepackage{ulem}
\usepackage{multirow}

\usepackage{physics}
\providecommand{\dif}{\mathrm{d}} \def\d{\dif}
\setstcolor{red}

\newcommand{\beq}{\begin{equation}}
\newcommand{\eeq}{\end{equation}}
\newcommand{\bea}{\begin{eqnarray}}
\newcommand{\eea}{\end{eqnarray}}
\newcommand{\non}{\nonumber}
\providecommand{\dif}{\mathrm{d}} 

\def\RS{\Sigma}

\def\d{\dif}

\begin{document}

\title{Circular equatorial orbits of extended bodies with spin-induced quadrupole around a Kerr black hole: Comparing spin-supplementary conditions}

\author{Misbah Shahzadi
\orcidlink{0000-0002-3130-1602}}
\email{misbahshahzadi51@gmail.com}
\affiliation{Astronomical Institute of the Czech Academy of Sciences, Bo\v{c}n\'{i} II 1401/1a, CZ-141 00 Prague, Czech Republic}

\author{Georgios Lukes-Gerakopoulos
\orcidlink{0000-0002-6333-3094}}
\email{gglukes@gmail.com}
\affiliation{Astronomical Institute of the Czech Academy of Sciences, Bo\v{c}n\'{i} II 1401/1a, CZ-141 00 Prague, Czech Republic}

\author{Martin Kolo{\v s}
\orcidlink{0000-0002-4900-5537}}
\email{martin.kolos@physics.slu.cz}
\affiliation{Research Centre for Theoretical Physics and Astrophysics, Institute of Physics, \\Silesian University in Opava, Bezru\v{c}ovo n\'{a}m.13, CZ-74601 Opava, Czech Republic}

\date{\today}
\begin{abstract}
The worldline of an extended body in curved spacetime can be described by the Mathisson-Papapetrou-Dixon equations when its centroid, i.e., its center of mass, is fixed by a spin supplementary condition (SSC). Different SSC choices result in distinct worldlines. To examine the properties of these choices, we investigate the frequency of circular equatorial orbits of extended bodies within the pole-dipole-(spin-induced) quadrupole approximation moving around a Kerr black hole (BH) for the Tulzcyjew-Dixon (TD) and the Mathisson-Pirani (MP) SSCs. First, we examine similarities and discrepancies in the prograde and retrograde orbital frequencies by expanding these frequencies in power series of the spin without taking into account the fact that both the position of the centroid and the spin measure change under the transition from one SSC to another. Then, by taking into account the centroid transition laws, we examine the orbital frequencies convergence between the non-helical MP frame to the TD frame. In particular, we demonstrate that, in analogy to the pole-dipole approximation, the transition from one circular orbit to another within the pole-dipole-(spin-induced)quadrupole approximation under a change in the SSC, results in convergence between the SSCs only up to certain terms in the spin expansion and does not extend to the entire power series. Finally, we discuss the innermost stable circular orbits (ISCOs) in the pole-dipole-(spin-induced)quadrupole approximation under TD and MP SSCs.
\end{abstract}

 \maketitle
\setcounter{tocdepth}{2}

\section{Introduction} \label{sec:Intro}

The description of gravitational interaction between the components of a binary system within the framework of general relativity requires careful consideration of their internal structure. The orbital dynamics of two bound compact objects has been extensively studied in the literature through a variety of approximation techniques. These techniques span from post-Newtonian approximation \cite{Blanchet:2014LRR:}, which can be enhanced with effective field theory techniques \cite{Levi:2020RPPh}, to gravitational self-force corrections to the geodesic motion \cite{Poisson-etal:2011LRR:}. All these approaches can be exploited to produce a unified phenomenological model to cover all possible binary configurations in the ``effective-one-body'' (EOB) framework \cite{Buonanno-Damour:1999PhRvD:,Albertini2022PhRvD,Albertini2022PhRvDI,Albertini2022PhRvDII,Albertini2024PhRvD}.

When one extended body in a binary system has significantly smaller mass than the other, then the system can be simplified to the dynamics of an extended body in a fixed background gravitational field generated by the more massive object. Within this approximation, a self-consistent framework for the evolution of pole-dipole sources was developed by Mathisson \cite{Mathisson:1937:Acta:}, and Papapetrou \cite{Papapetrou:1951:PPSA:}, and later generalized to bodies endowed with higher multipoles by Dixon \cite{Dixon:1970:RSPSA:}, known now as the Mathisson-Papapetrou-Dixon (MPD) model. This formulation describes the motion of an extended test body with internal degrees of freedom in a fixed background gravitational field, neglecting gravitational radiation reaction. 

The motion of an extended test body deviates from the geodesic trajectory of a test particle due to the influence of forces coupling the background's curvature with the body's multipoles. A striking difference is that, though the geodesic motion on a Kerr BH background corresponds to an integrable system, in their pioneering work, Suzuki and Maeda \cite{Suzuki-Maeda:1997:prd:} showed that the pole-dipole body motion around a Schwarzschild BH has chaotic behavior. Ref.~\cite{Tod-etal:1976:NCimB:} explored the effects of the secondary spin on the motion of a pole-dipole body on an equatorial plane around a BH, while Ref.~\cite{Rau-etal:2016:PhRvD:} provided a study of generic orbits of pole-dipole bodies around a BH in the frequency domain. In general, the influence of the secondary spin on the orbital motion around BHs within the pole-dipole approximation has been extensively studied by several authors \cite{Semerak:1999:MNRAS:,Kyrian-Semerak:2007:MNRAS:,Eva-et-al:2014:PhRvD,Gera-etal:2014:PhRvD:,Gera-etal:2016:PhRvD:,Filipe-etal:2018:PhRvD:,Zelenka-etal:2020:PhRvD:,Ias:Gera:The:2021:PhRvD:,Skoupy2025PhRvL,Sha-etal:2021:EPJC:}. 

The natural step to go beyond the pole-dipole approximation is to include the quadrupole moment to the MPD equations and to investigate the dynamics of extended bodies within the pole-dipole-quadrupole approximation \cite{Ehlers-Rudolph:1977:GReGr:,Costa2015emrg}. Steinhoff and Puetzfeld \cite{Steinhoff-Puetzfeld:2010PhRvD:,Steinhoff-Puetzfeld:2012PhRvD:} developed a general framework within such an MPD model to incorporate both quadratic spin corrections and tidal interactions, with a particular focus on analyzing the binding energy of the system derived from the associated effective potential, suggesting a way to find circular equatorial orbits. Bini and Geralico \cite{Bini-Geralico:2014PhRvD:} analyzed the equatorial motion of extended test bodies with a general quadrupole tensor, while Bini et al. \cite{Bini-etal:2013PhRvD:,Bini-etal:2015PhRvD:} studied the influence of spin-induced quadrupoles in the equatorial motion around Schwarzschild and Kerr BHs; determining the radial effective potential, they were able to numerically calculate the position of the ISCO. Timogiannis et al. \cite{Tim-etal:2023PhRvD:} formulated the momentum-velocity relation for the pole-diple-quadrupole body under the MP SSC and calculated analytically the orbital frequency for the circular equatorial orbits in the Kerr background. The influence of spin-induced quadrupole moment on orbital motion around BHs and neutron stars has been studied by several authors \cite{Bini-Andrea:2015PhRvD:,Han-Cheng:2017GReGr:,Li-etal:2022MNRAS:,Mukherjee-etal:2022GReGr:,Zhang-Jiang:2022PhLB:,Rahman-Bhattacharyya:2023PhRvD:,Sang-etal:2023EPJC:}, but not so extensively as the pole-dipole case.

Regarding the formalism of the MPD equations, Bailey and Israel \cite{Bailey-Israel:1975CMaPh:} demonstrated that the MPD equations can be derived from an effective Lagrangian. Several Hamiltonian functions providing the MPD equations in the pole-dipole approximation have been suggested over the years \cite{Barausse-etal:2009PhRvD:,Witzany:2019CQGra,Witzany:2019PhRv,Ramond2025PhRvD}, but only a few considered the pole-dipole-quadrupole case \cite{Vines:2016PhRvD}. The fact that MPD equations can be derived from a Hamiltonian function implies that they possess a symplectic structure. This is useful for studying numerically the MPD equations, which entail a bunch of interesting numerical challenges. The efficient integration of equations of motion over a long time interval needs the structure preserving algorithms, i.e., symplectic schemes, which have been successfully applied for simulations in various fields of general relativity \cite{Hairer:2006:,Shahzadi-etal:2023PhRvD:} and in particular for the MPD case \cite{Gera-etal:2014:PhRvD:}.

In this paper, we investigate the dynamics of an extended test body in the pole-dipole-(spin-induced)quadrupole approximation moving in a Kerr spacetime, under the TD SSC \cite{Tulczyjew59,Dixon:1964:NCim:} and the MP SSC \cite{Mathisson:1937:Acta:,Pirani56}. The motion is restricted to the equatorial plane, with the spin vector of an extended body aligned with the rotational axis of the central object. 
Following the work in Refs.~\cite{Ias:Gera:The:2021:PhRvD:,Timogiannis-etal:2022PhRvD:}, where the orbital frequencies under different SSCs were expanded in powers of the spin measure, and the transition of a pole-dipole body from one centroid to another centroid of the same body was studied, we do the same thing in the pole-dipole-(spin-induced)quadrupole approximation. Specifically, we examine up to which order in spin a new centroid preserves the orbital frequency of the original centroid as the SSC is changed. We adopt the same form of the mass quadrupole tensor as used in \cite{Steinhoff-Puetzfeld:2012PhRvD:}, while ignoring the quadrupolar tidal effects. In general, the quadrupole tensor shares the same symmetries as the Riemann tensor and is completely characterized by two symmetric, trace-free spatial tensors, i.e., the mass quadrupole (electric) and the current quadrupole (magnetic) tensors, whose role has been investigated in previous works \cite{Bini-Geralico:2014PhRvD:}. Here, we consider the rotational deformation induced by a mass quadrupole tensor, with a constant proportionality parameter which may be regarded as the polarizability of the object. For neutron stars, this parameter depends on the equation of state, while for BHs, it is exactly 1. We treat it as a free parameter of the model, as it can influence the associated observables as the orbital frequency. 

As in the case of pole-dipole, where the question is whether one can measure the secondary spin through its impact on gravitational waves emitted by compact object binary inspiraling systems with extreme-mass ratio known as extreme-mass ratio inspirals (EMRIs) \cite{Piovano:2021PhRvD}; similarly, there is an ongoing effort to assess the detectability of the imprint of the spin-induced quadrupole term in the gravitational waves from EMRIs \cite{Rahman-Bhattacharyya:2023PhRvD:}. Our contribution to such discussions is to assess whether the choice of the center of mass can play any role in the detection of the spin and quadrupole of the extended body. We approach this by studying circular equatorial orbital frequencies of pole-dipole-(spin-induced) quadrupole bodies moving around a Kerr BH.

The structure of this article is as follows: Sec.~\ref{sec:MPD} introduces the formalism of the MPD model within the pole-dipole-quadrupole approximation under both the TD and MP SSCs. Sec.~\ref{sec:SBKerr} examines the equatorial circular orbits of an extended test body with a spin-induced quadrupole structure moving in a Kerr background, under both the TD and MP SSCs. Sec.~\ref{sec:SSCcomp} presents a comparison of the orbital frequencies of extended test bodies across different SSCs, along with the corresponding shifts between these SSCs. Finally, Sec.~\ref{sec:Conc} summarizes the key findings of this work.  

We use the geometric units, where the speed of light and the gravitational constant are set to $c = G =1$. The Riemann tensor is defined as: $R^{\rho}_{\ \sigma\mu\nu} = \partial_\mu \Gamma^{\rho}_{\nu\sigma} - \partial_\nu \Gamma^{\rho}_{\mu\sigma} + \Gamma^{\rho}_{\mu\lambda} \Gamma^{\lambda}_{\nu\sigma} - \Gamma^{\rho}_{\nu\lambda} \Gamma^{\lambda}_{\mu\sigma}
 $, while the Christoffel symbols are calculated using a metric with the signature $(-,+,+,+)$. Greek indices run from 0 to 3.
The central BH's mass is represented by $M$,  while for the test body, we use two notions of mass: $\mu=\sqrt{-p^\alpha p_\alpha}$, and $m=-u^\alpha p_\alpha$. Dimensionless quantities are represented by a symbol $\hat{}$. For instance, the dimensionless radius $r$ is expressed as $\hat{r}=r/M$ and the dimensionless Kerr parameter as $\hat{a}=a/M$. However, for the dimensionless spin parameter, we use different conventions depending on the choice of the SSCs. For the TD SSC, the dimensionless spin measure is given by $\sigma=\frac{S}{\mu M}$, whereas for the MP SSC, $\sigma=\frac{S}{m M}$. Similarly, the dimensionless polarizability parameter is denoted as $\hat{k}=\mu k$ for the TD SSC, and $\hat{k}=m k$ for the MP SSC.

\section{Mathisson-Papapetrou-Dixon equations} \label{sec:MPD}

The derivation of the MPD equations follows from the conservation of the energy-momentum tensor $T^{\alpha \beta}$, which characterizes the body. These equations incorporate contributions from the monopole, dipole, quadrupole, or higher-ordered moments of the stress-energy tensor, which are calculated by integrating the stress-energy tensor over a three-dimensional spacelike hypersurface, defined by a constant but arbitrary coordinate time $t$. In this formalism, the zeroth multipole moment, often referred to as the mass-monopole can be encoded in the four-momentum $p^{\alpha}$, while the first multipole moment, called the spin-dipole, in the anti-symmetric spin tensor $S^{\alpha \beta}$. The quadrupole moment is encoded in the rank four quadrupole tensor $J^{\alpha \beta \gamma \delta}$. In practice, the multipole expansion is truncated at a specific order; in our study, this truncation occurs at the quadrupole, neglecting all the higher-ordered multipoles. Hence, this pole-dipole-quadrupole approximated body is characterized by its mass, its spin, and a quadrupole tensor. In the MPD framework it is assumed that the extended body is a test body; hence, it does not deform the background spacetime due to its own gravity. In the above described approximation, the MPD equations of motion read
\bea \label{eq:spin1}
    \dot{x}^\alpha &=& u^\alpha,\\\label{eq:spin2}
    \dot{p}^\alpha &=& -\frac{1}{2} R^{\alpha}_{\mu\nu\rho} S^{\nu\rho} u^\mu - \frac{1}{6} J^{\mu \nu \rho \sigma} \nabla^{\alpha} R_{\mu \nu \rho \sigma} ,\\\label{eq:spin3}
    \dot{S}^{\alpha\beta} &=& p^\alpha u^\beta - u^\alpha p^\beta + \frac{4}{3} J^{\mu \nu \rho [\alpha} R^{\beta]}_{\,\, \rho \mu \nu},
\eea
where $\dot{~}$ denotes the covariant derivative with respect to the proper time $\tau$, $R^{\alpha}_{\mu\nu\rho}$ defines the Riemann curvature tensor, $S^{\alpha\beta}$ is the antisymmetric spin tensor, $u^\alpha$ is the four-velocity\footnote{Since we have chosen the proper time as our affine parameter, we have introduced the constraint $u^\mu u_\mu=-1$ in our MPD framework.}, $p^\alpha$ is the four-momentum, and $J^{\alpha \beta \gamma \delta}$ is the quadrupole tensor with the following symmetries
\bea 
J^{\alpha \beta \gamma \delta} &=& J^{[\alpha \beta] [\gamma \delta]}= J^{\gamma \delta \alpha \beta},\\\
J^{[\alpha \beta \gamma] \delta} &=& 0 \iff J^{\alpha \beta \gamma \delta} + J^{\beta \gamma \alpha \delta} + J^{\gamma \alpha \beta \delta} = 0.  
\eea
Thus, the quadrupole tensor $J^{\alpha \beta \gamma \delta}$ has the same algebraic symmetries as the Riemann tensor and is solely defined by the internal matter structure of the body \cite{Steinhoff-Dirk:2012:PhRvD:}. 

The right-hand side of Eq.~\eqref{eq:spin2} indicates the spin-orbit coupling through a strong gravitational field. For $J^{\alpha \beta \gamma \delta}=0$, the MPD equations \eqref{eq:spin1}-\eqref{eq:spin3} reduce to pole-dipole approximation. Furthermore, in the absence of the spin ($S^{\alpha \beta} =0$), we recover the geodesic equation $\dot{p}^\alpha= 0$, i.e., a non-spinning body follows the geodesic motion. However, in general, an extended spinning test body does not follow a geodesic path \cite{Mukherjee-etal:2022GReGr:} and the reference worldline deviates from the geodesic motion due to the influences of the spin-curvature force (coupling). It is important to note that MPD equations describe only the evolution of the momentum and of the spin of the body, whereas the quadrupole moment is determined by the matter distribution of the body.

The MPD equations are the first-order non-linear ordinary differential equations, but not a closed set. This means that there are fewer equations than necessary to evolve the system. In order to address this issue, additional conditions are required to determine a reference point about which the spin and other higher multipole moments of the body can be calculated. The center of mass of the body can serve as this reference point. In relativity, however, such a point is observer-dependent \cite{Costa2015emrg}. Various constraining conditions have been proposed to complete the set of MPD equations, which are known as SSCs \cite{Semerak:1999:MNRAS:}. An SSC fixes the center of mass, often called the centroid, and is described by the general relation
\beq
V^\alpha S_{\alpha \beta} = 0,
\eeq
where $V^{\alpha}$ is a future-oriented timelike vector, typically subjected to the normalization condition $V^\alpha V_\alpha = -1$. The spin vector in terms of reference vector $V^\alpha$ is defined as
\beq\label{eq:spin-vector}
S_\alpha = -\frac{1}{2}\eta_{\alpha\beta\mu\nu} V^{\beta} S^{\mu\nu},
\eeq
and the inverse relation of Eq.~\eqref{eq:spin-vector} is given by
\beq
 S^{\alpha\beta} = -\eta^{\alpha\beta\gamma\delta} S_{\gamma} V_{\delta},\label{eq:spin-tensor}
\eeq
where $\eta_{\alpha\beta\mu\nu} = \sqrt{-g} ~ \epsilon_{\alpha\beta\mu\nu}$ is the Levi-Civita tensor, $\epsilon_{\alpha\beta\mu\nu}$ denotes the Levi-Civita symbol, and $g$ is the determinant of the background spacetime. After performing some calculations, Eq.~\eqref{eq:spin-tensor} implies that
\beq\label{eq:spinTen2SpinV}
S^{\alpha \beta} S_{\beta \mu} = S^{\alpha}_{\mu} - S^2 h^{\alpha}_{\mu},
\eeq
where $h^{\alpha}_{\mu} = \delta^{\alpha}_{\mu} + V^{\alpha} V_{\mu}$ is the space projector orthogonal to $V^{\alpha}$, and $S$ is the magnitude of the spin, given by
\beq\label{eq:spin_magnitude}
S^2 = S^\alpha S_\alpha = \frac{1}{2} S^{\alpha\beta} S_{\alpha\beta}.
\eeq
The quadrupole moment can be decomposed using a reference vector in the following way \cite{Harte:2020PhRvD:}
\beq\label{eq:Jquadrupole}
J^{\mu \nu \rho \sigma} = \tau^{\mu \nu \rho \sigma} - 3 V^{[\mu} Q^{\nu][\rho} V^{\sigma ]} - V^{[\mu} \Pi^{\nu] \rho \sigma} - V^{[\rho} \Pi^{\sigma] \mu \nu},
\eeq
where $\tau^{\mu \nu \rho \sigma}$ is the stress quadrupole, $Q^{\alpha \beta}$ is the mass quadrupole, and $\Pi^{\alpha \beta \gamma}$ is the flow quadrupole \cite{Ehlers-Rudolph:1977:GReGr:}. In this study, we concentrate on the spin-induced quadrupole model, where the mass quadrupole is expressed as
\beq \label{eq:SpinQuad}
Q^{\alpha \beta} = k\, S^{\alpha}_{\gamma} S^{\beta \gamma}. 
\eeq
Here, $k$ is the polarizability constant, whose value depends on the equation of state of the object. It is normalized such that $\hat{k}=1$ corresponds to a BH. For neutron stars, the value of $\hat{k}$ typically ranges between 3.1 and 7.4 \cite{William-Poisson:1999:ApJ:}. More recent work found values between 5 and 6 \cite{Urbanec-etal:2013MNRAS:}.

In analogy to the pole-dipole case ($J^{\alpha \beta \gamma \delta}=0$), the following quantity
\beq\label{eq:killing-field1}
    C({\xi}) = p^{\alpha} \xi_{\alpha} -\frac{1}{2} S^{\alpha\beta} \xi_{\alpha ; \beta},
\eeq
remains conserved in the pole-dipole-quadrupole approximation if $\xi^\alpha$ is a Killing vector, satisfying the condition $\nabla_{(\beta}\, \xi_{\alpha)} = 0$. Moreover, the conserved nature of Eq.~\eqref{eq:killing-field1} holds even at all higher multipole orders. The existence of other conserved quantities depends on the choice of SSCs. For the spin length $S$, by contracting Eq.~\eqref{eq:spin3} with $S_{\alpha\beta}$ and using the Eq.~\eqref{eq:spin_magnitude}, one can obtain \cite{Steinhoff-Dirk:2012:PhRvD:}
\beq\label{eq:spin-evolution}
S \dot{S} = S_{\alpha\beta}\, p^{\alpha} u^{\beta} + \frac{2}{3} S_{\alpha \beta} J^{\mu \nu \rho \alpha} R^{\beta}_{\rho \mu \nu} ,
\eeq
for the mass $\mu^2=-p^\nu p_\nu$, by contracting Eq.~\eqref{eq:spin3} with $\dot{p}_{\alpha} p_{\beta}$ and contracting Eq.~\eqref{eq:spin2} with $u_a$ we arrive to the following relation 
\bea\non \label{eq:DynMassEvol}
\dot{\mu} &=&  \frac{\mu}{6m}  J^{\alpha \beta \gamma \eta} \, \dot{R}_{\alpha \beta \gamma \eta}- \frac{4}{3 \mu m} J^{\mu\nu\rho[\alpha} R^{\beta]}_{\rho\mu\nu}  \, \dot{p}_{\alpha} p_{\beta} \\\ &+& \frac{1}{\mu m} \dot{S}^{\alpha \beta}\, \dot{p}_{\alpha} p_{\beta}, 
\eea
while for the mass $m=-u^\nu p_\nu$, by contracting Eq.~\eqref{eq:spin3} with $\dot{u}_{\alpha} u_{\beta}$ and contracting Eq.~\eqref{eq:spin2} with $u_a$ we arrive to the following relation 
\bea\non \label{eq:KinMassEvol}
\dot{m} &=&  \frac{1}{6}  J^{\alpha \beta \gamma \eta} \, \dot{R}_{\alpha \beta \gamma \eta}+ \frac{4}{3} J^{\mu\nu\rho[\alpha} R^{\beta]}_{\rho\mu\nu}  \, \dot{u}_{\alpha} u_{\beta} \\\ &+& \dot{S}^{\alpha \beta}\, \dot{u}_{\alpha} u_{\beta}. 
\eea

\subsection{SSCs} \label{sec:SSC}

In this work, we only focus on the TD and the MP SSCs, which are described in the following sections.

\subsubsection{TD SSC} \label{sec:TD}

The TD SSC defines a unique future-oriented timelike vector as $V^{\alpha} = p^{\alpha} / \mu$. This SSC is given by \cite{Dixon:1970:RSPSA:,Dixon:1974:RSPTA:}
\beq\label{eq:TDSSC}
    p_{\mu} S^{\mu \nu} = 0,
\eeq
and specifies a unique worldline. The TD SSC is extensively used in numerical calculations, mostly due to the existence of an explicit relation between the four-velocity and the four-momentum of the spinning test body
\beq\label{eq:4v4mTD}
    u^\mu = \mathcal{P}^{\mu} + w^\mu,
\eeq
where $\mathcal{P}^\mu$ and $w^\mu$ are defined as
\bea\non
\mathcal{P}^\mu &=& \frac{1}{\mu^2} \left[m p^{\mu} - F_\alpha S^{\mu \alpha} - p_{\alpha} F^{\mu \alpha} \right],\\\
w^\mu &=&  \frac{2 S^{\mu\nu} R_{\nu\lambda\rho\sigma}  S^{\rho\sigma}}{4 \mu^{2} + S^{\alpha\beta} S^{\gamma\delta} R_{\alpha\beta\gamma\delta}}  \mathcal{P}^{\lambda}.
\eea
Here, we have introduced 
\bea
F^{\alpha} &=& - \frac{1}{6} J^{\mu \nu \rho \sigma} \nabla^{\alpha} R_{\mu \nu \rho \sigma}, \\
F^{\alpha \beta} &=& \frac{4}{3} J^{\mu \nu \rho [\alpha} R^{\beta]}_{\,\, \rho \mu \nu}.
\eea
Using the normalization condition of four-velocity, i.e., $u^\alpha u_\alpha = -1 $, we can find as in Ref.~\cite{Tim-etal:2023PhRvD:}  that the kinematical mass $m$ of spin-induced spinning body, in terms of dynamical mass $\mu$, is given by 
\beq\label{eq:m}
m_\pm = \frac{-\mathcal{B} \pm \sqrt{\mathcal{B}^2  - 4 \mathcal{A}\, \mathcal{C}}}{2 \mathcal{A}},
\eeq
where 
\bea \label{eq:ABC}
\mathcal{A} &=& \frac{\mathcal{F}_4}{\mathcal{F}_0^2} - \frac{1}{\mu^2}, \quad \mathcal{B} = \frac{\mathcal{F}_2}{\mathcal{F}_{0}} + \frac{\mathcal{F}_5}{\mathcal{F}_{0}^2},  \\
\mathcal{C} &=& 1 + \mathcal{F}_1 + \frac{\mathcal{F}_3}{\mathcal{F}_0} + \frac{\mathcal{F}_6}{\mathcal{F}_0^2},
\eea
and
\bea
\mathcal{F}_0 &=& 4 \mu^{2} + S^{\alpha\beta} S^{\gamma\delta} R_{\alpha\beta\gamma\delta},\\\non
\mu^4 \mathcal{F}_1  &=& F^\alpha S^{\mu k} F_k S_{\mu \alpha}  + 2 p^\nu S^{\mu k} F_k F_{\mu \nu}  \\\ &+& p^\lambda F^{\mu \nu} F_{\mu \lambda} p_{\nu},\\
\mu^4 \mathcal{F}_2  &=& - 4 R_{k \lambda \nu \sigma} S^{\nu \sigma} S^{\mu k} p^{\lambda} \left(S_{\mu \alpha} F^\alpha + F_{\mu \beta} p^\beta \right), \\\non
\mu^4 \mathcal{F}_3  &=& 4 R_{k \lambda \nu \sigma} S^{\nu \sigma} S^{\mu k} \left( S_{\mu \alpha} F^\alpha + F_{\mu \beta} p^\beta \right) \\ &\times& \left( F_{\gamma} S^{\lambda \gamma} + p_{\delta} F^{\lambda \delta} \right),\\
\mu^4 \mathcal{F}_4  &=& 4 R_{k \lambda \nu \sigma} S^{\nu \sigma} S^{\mu k} R^{\pi}_{\rho \epsilon \zeta} S^{\epsilon \zeta} S_{\mu \pi} p^\rho p^\lambda S^{\epsilon \zeta},\\\non
\mu^4 \mathcal{F}_5  &=& - 4 R_{k \lambda \nu \sigma} S^{\nu \sigma} S^{\mu k} R^{\pi}_{\rho \epsilon \zeta} S^{\epsilon \zeta} S_{\mu \pi} p^\lambda S^{\epsilon \zeta} \left(S^{\rho \alpha} F_{\alpha} \right. \\\non  &+& \left.  F^{\rho \beta} p_\beta \right) -
4 R_{k \lambda \nu \sigma} S^{\nu \sigma} S^{\mu k} R^{\pi}_{\rho \epsilon \zeta} S^{\epsilon \zeta} S_{\mu \pi} p^\rho S^{\epsilon \zeta} \\ &\times& \left(S^{\lambda \xi} F_\xi + F^{\lambda \tau} p_\tau \right),\\\non
\mu^4 \mathcal{F}_6  &=& - 4 R_{k \lambda \nu \sigma} S^{\nu \sigma} S^{\mu k} R^{\pi}_{\rho \epsilon \zeta} S^{\epsilon \zeta} S_{\mu \pi} S^{\rho \alpha} F_\alpha \left (S^{\lambda \xi} F_\xi \right. \\\non  &+& \left. F^{\lambda \tau} p_\tau \right) +
4 R_{k \lambda \nu \sigma} S^{\nu \sigma} S^{\mu k} R^{\pi}_{\rho \epsilon \zeta} S^{\epsilon \zeta} S_{\mu \pi} F^{\rho \beta} p_\beta \\ &\times& \left (S^{\lambda \xi} F_\xi + F^{\lambda \tau} p_\tau \right).
\eea
From the two solutions $m_{-}$ and $m_{+}$, we use $m_{-}$ in our work, as it is the physically accepted one. The reasoning is the following: in the pole-dipole limit, the mathematical expression for $\mathcal{A}$ does not change, but the other two quantities in Eq.~\eqref{eq:ABC} reduce to $\mathcal{B}=0$ and  $\mathcal{C}=1$. Hence, Eq.~\eqref{eq:m} in the pole-dipole limit leads to $m_\pm=\frac{\pm\sqrt{-\mathcal{A}}}{\mathcal{A}}$, implying that $\mathcal{A} < 0$ and leading to $m_->0$. Fixing the kinematical mass $m$ is the last necessary step that allows us to evaluate numerically the MPD equations under TD SSC.

It is clear from Eq.~\eqref{eq:4v4mTD} that the four-velocity $u^{\alpha}$ and the four-momentum $p^{\alpha}$ are, in general, not parallel for an extended body within the pole-dipole-quadrupole approximation under TD SCC \eqref{eq:TDSSC}. The four-velocity $u^\mu$ is enforced to be time-like ($u^\alpha u_\alpha = -1$) and the four-momentum is expected to be timelike as well, since $\mu^2 = - p^{\alpha} p_{\alpha} >0$ \cite{Han-Reg:1974:AnPhy:}.

The evolution~\eqref{eq:DynMassEvol} of the dynamical rest mass $\mu$ for TD SSC is reduced to 
\beq\label{eq:TD-mass-Evolution}
\dot{\mu} =  \frac{\mu}{6m}  J^{\alpha \beta \gamma \eta} \, \dot{R}_{\alpha \beta \gamma \eta}- \frac{4}{3 \mu m} J^{\mu\nu\rho[\alpha} R^{\beta]}_{\rho\mu\nu}\, \dot{p}_{\alpha} p_{\beta}, 
\eeq
while the spin length to
\beq\label{eq:TD-spin-Evolution}
S \dot{S} = \frac{2}{3} S_{\alpha \beta} J^{\mu \nu \rho \alpha} R^{\beta}_{\rho \mu \nu}.
\eeq
It can be seen from Eq.~\eqref{eq:TD-mass-Evolution} that, in contrast to the pole-dipole approximation, the mass $\mu$ and spin magnitude $S$ are not constants of motion for the pole-dipole-quadrupole approximation. However, the spin evolution remains conserved under the conditions provided in \cite{Steinhoff-Dirk:2012:PhRvD:}
\beq
\dot{S} = 0,
\eeq
which includes the spin-induced quadrupole case. Finally, the mass $\mu$ remains conserved only for circular equatorial orbits around BHs \cite{Tim-etal:2023PhRvD:}, but not for the eccentric equatorial motion.

\subsubsection{MP SSC} \label{sec:MP}

The MP SSC is characterized by a reference timelike four-vector which coincides with the four-velocity $u^\alpha$ of the spinning body, i.e., \( V^\alpha = u^\alpha \). As a result, the observer's frame of reference aligns with the rest frame of the body, meaning that the observer is co-moving with the body. The MP SSC is mathematically expressed in the form 
\beq \label{eq:MPSSC}
u^\alpha S_{\alpha \beta}=0.
\eeq
Note that the MP SSC does not fix a unique world line \cite{Filipe-etal:2018:PhRvD:} in contrast to TD SSC. In general, an MP SSC choice leads to helical motion, which was considered for a long time to be unphysical, but, in fact, it is just a poor selection of the center of mass \cite{Costa2015emrg}. In this work, the MP SSC is chosen in such a way that it provides non-helical motion.

Under MP SSC, the evolution of the kinematical mass~\eqref{eq:KinMassEvol} reduces to
\beq\label{eq:MP-mass-Evolution}
\dot{m} =  \frac{1}{6}  J^{\alpha \beta \gamma \eta} \, \dot{R}_{\alpha \beta \gamma \eta}+ \frac{4}{3} J^{\mu\nu\rho[\alpha} R^{\beta]}_{\rho\mu\nu}  \, \dot{u}_{\alpha} u_{\beta} .
\eeq
Equation \eqref{eq:MP-mass-Evolution} demonstrates that, unlike in the pole-dipole approximation, the kinematical rest mass \( m \) is not a constant of motion within the pole-dipole-quadrupole approximation. However, the kinematical rest mass is constant for the case of circular equatorial orbits around BHs\footnote{But, not for the eccentric equatorial orbits, which have implications on finding the inner stable circular equatorial orbit}.

Substituting Eqs.~(\ref{eq:Jquadrupole}) and (\ref{eq:MPSSC}) into Eq.~(\ref{eq:spin-evolution}), we obtain
\beq\label{eq:spin_MP1}
S \dot{S} = - S^{\alpha\beta} R_{\alpha\gamma\delta\epsilon} u^\gamma u^\epsilon Q_{~\beta}^{\delta} = - X_{\alpha\delta} Y^{\alpha\delta},
\eeq
where $X_{\alpha\delta}= R_{\alpha\gamma\delta\epsilon} u^\gamma u^\epsilon$ is a symmetric tensor and using Eq.~\eqref{eq:SpinQuad}, we define
\beq
Y^{\alpha\delta} = S^{\alpha\beta} Q_{~\beta}^{\delta}=k S^{\alpha\beta} S_{\beta\gamma} S^{\gamma\delta},
\eeq
which is an antisymmetric tensor due to the antisymmetric property of the spin-tensor $S^{\alpha \beta}$. Using the fact that the contraction of symmetric and antisymmetric tensors vanishes, Eq.~(\ref{eq:spin_MP1}) turns out to be
\beq
\dot{S} = 0.
\eeq
It shows that the spin evolution of the pole-dipole-(spin-induced) quadrupole body remains conserved under MP SSC.

\section{Spinning bodies in Kerr background} \label{sec:SBKerr}

We consider the motion of extended test bodies around a Kerr BH of mass $M$ and the corresponding geometry is characterized by the line element
\beq
\d s^2 = g_{tt}\d{t}^2 + g_{rr}\d{r}^2 + g_{\theta\theta}\d\theta^2 + g_{\phi\phi}\d\phi^2 +2g_{t\phi}\d{t}\d\phi, \label{eq:Kerr-MOGMetric}
\eeq
with the nonzero components of the metric tensor $g_{\mu\nu}$ taking in the standard Boyer-Lindquist coordinates the form
\bea
g_{tt} &=& -\left(\frac{\Delta-a^{2}\sin^{2}\theta}{\RS}\right), \quad
g_{rr} = \frac{\RS}{\Delta}, \quad g_{\theta\theta} = \RS,\nonumber\\
g_{\phi\phi} &=& \frac{\sin^{2}\theta}{\RS}\left[(r^{2}+a^{2})^{2}-\Delta
a^{2}\sin^{2}\theta\right], \nonumber\\
g_{t\phi}&=&\frac{a\sin^{2}\theta}{\RS}\left[\Delta-(r^{2}+a^{2})\right] ,
\label{eq:MetricCoef}
\eea
where
\bea
\Delta &=& r^2 - 2GMr + a^2, \nonumber\\ \RS &=& r^2 + a^2 \cos^2\theta.
\eea
The outer horizon of the BH is situated at
\beq
r_{+} = M + \sqrt{ M^2 -a^2 }. \label{eq:horizon}
\eeq

\subsection{Equatorial circular orbits}\label{sec:CEOs} 

We consider the motion of spinning test bodies on an equatorial plane, i.e., $\theta = \frac{\pi}{2},~p^{\theta}=0$. As discussed in detail in Refs.~\cite{Steinhoff-Dirk:2012:PhRvD:,Tim-etal:2023PhRvD:}, to constrain the body under TD and MP SSCs on the equatorial plane, its spin-vector has to  be orthogonal to the orbital plane, i.e., 
\beq\label{eq:spin-vec}
  S^{\alpha} = S^{\theta} \delta_{\theta}^{\alpha}.
\eeq
Using Eq.~\eqref{eq:spin_magnitude} and Eq.~\eqref{eq:spin-vec}, the spin-vector can be expressed in terms of spin magnitude $\mathcal{S}$ as
\beq\label{eq:polar-comp}
    S_{\theta} = - \sqrt{g_{\theta \theta}} ~ \mathcal{S},
\eeq
with $\mathcal{S}>0$ ($\mathcal{S}<0$) corresponding to a spin-vector (anti-) aligned with the total angular momentum, which by convention is always pointing along the positive z-direction. Taking into account for circular orbits $p^r=u^r=0$, the non-zero components of spin-tensor calculated using Eqs.~\eqref{eq:spin-tensor} and \eqref{eq:polar-comp} take the form
\bea\label{eq:ST-1}
	S^{tr}&=& -\mathcal{S} \sqrt{-\frac{g_{\theta \theta}}{g}} ~ V_{\phi} = -S^{rt},\\\label{eq:ST-2}
	S^{r\phi}&=&- \mathcal{S} \sqrt{-\frac{g_{\theta \theta}}{g}}~V_{t} = -S^{\phi r}.
\eea
The orbital frequency of a spinning test body on a circular orbit can be written in the form
\beq\label{eq:def_omega}
\Omega = \frac{u^\phi}{u^t}.
\eeq
Taking into account that the radial and the polar components of the four-velocity are zero ($u^r = u^\theta = 0$), the normalization condition of four-velocity, i.e., $u^\alpha u_\alpha = -1$, leads to
\beq\label{eq:utOmega}
u^t = \frac{1}{\sqrt{- g_{tt} - 2 g_{t \phi} \Omega - g_{\phi \phi } \Omega^2}}.
\eeq

The stationary and axisymmetry of the Kerr spacetime lead to the conservation of the energy $E$ and the $z$-component of the total angular momentum $J_z$ of the spinning test body, respectively. For equatorial circular orbits, these quantities are given by 
\bea\label{eq:energy}
E &=& - p_t + \frac{S}{2}\sqrt{-\frac{g_{\theta \theta}}{g}} \left(g_{t \phi,r} V_t - g_{tt,r} V_{\phi} \right),\\\label{eq:Jz}
J_z &=&  p_\phi + \frac{S}{2}\sqrt{-\frac{g_{\theta \theta}}{g}} \left(g_{t \phi,r} V_\phi - g_{\phi \phi,r} V_{t} \right).
\eea

\subsubsection{Equatorial circular orbits under TD SSC}\label{sec:ECO_TD} 

For circular equatorial orbits, the system of MPD Eqs.~\eqref{eq:spin2} and \eqref{eq:spin3} results in trivial identities, except for the components $\dot{p}^r$ and $\dot{S}^{t \phi}$. For these components, it holds that 
\beq\label{eq:nonZero_components}
\frac{\d p^r}{\d \tau} = \frac{\d S^{t \phi}}{\d \tau}=0.
\eeq
To solve the system of equations \eqref{eq:nonZero_components}, Ref.~\cite{Tim-etal:2023PhRvD:} introduced a quantity $W=p^{\phi}/p^t$, which through the definition of dynamical rest mass $\mu = \sqrt{-p^\alpha p_\alpha}$, establishes a relationship between the $p^t$ component and $W$ given by
\beq\label{eq:ptW}
p^t =  \frac{\mu}{\sqrt{-g_{tt} - 2g_{t \phi} W - g_{\phi \phi} W^2}}.
\eeq
Substituting Eqs.~\eqref{eq:utOmega} and \eqref{eq:ptW} into Eq.~\eqref{eq:nonZero_components} and after some calculations, we derive the equations in terms of $\Omega$ and $W$, expressed as follows 
\begin{widetext}
\bea\non
\frac{2 r^3 \left[ W \Omega  \left(\mu  r^5-3 a^3 M S\right) + M r^2 (S (2 W+\Omega )-\mu ) + a M \left(3 S \left(a (W+\Omega ) - r^2 W \Omega -1\right)+\mu  r^2 (W +\Omega -a W \Omega )\right) \right]}{\sqrt{r -r W^2 \left(a^2+r^2\right)-2 M (a W-1)^2}\, \sqrt{r -r \Omega ^2 \left(a^2+r^2\right)-2 M (a \Omega -1)^2}}\\\label{eq:omega_TD} + \frac{3 k M r S^2 \left(-10 a^3 W+5 a^2+W^2 \left(5 a^4+a^2 r (7 r-2 M)+2 r^4\right)+4 a r W (M-2 r)-2 M r+r^2\right)}{r \left(W^2 \left(a^2+r^2\right)-1\right)+2 M (a W-1)^2} =0,
\eea    
\bea\label{eq:W_TD}
\frac{3 k M r S^2 (a W-1) \left(W \left(a^2+r^2\right)-a\right)}{r \left(W^2 \left(a^2+r^2\right)-1\right)+2 M (a W-1)^2} = \frac{- r^3 \left( r^3 (\mu  \Omega +S W \Omega - \mu  W) - M S (a W-1) (a \Omega -1) \right)}{\sqrt{r-r W^2 \left(a^2+r^2\right)-2 M (a W-1)^2} \sqrt{r-r \Omega ^2 \left(a^2+r^2\right)-2 M (a \Omega -1)^2}} .
\eea
\end{widetext}

\begin{figure}
\begin{center}
\includegraphics[width=\hsize]{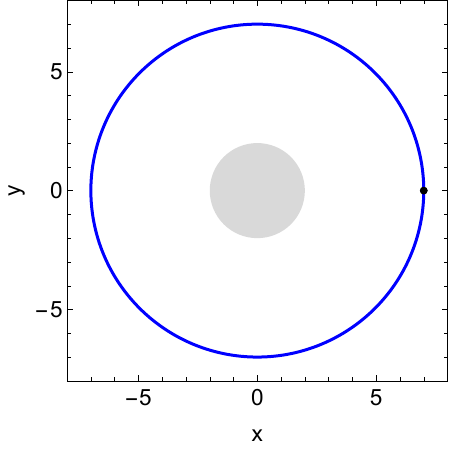}
\end{center}
\caption{An example of circular ($\hat{r} =7$, $p^r =0$) equatorial  ($\theta_0 = \pi/2$, $p^\theta=0$) orbit for a pole-dipole-(spin-induced)quadrupole under the TD SSC in Kerr background with $\hat{k}=1$, $\sigma=0.5$, $\hat{a}=0.9$, $\hat{E}=0.9316$, and $\hat{J}_z = 3.3569$, where ${\rm x}=\hat{r}\cos{\phi} \sin{\theta},~{\rm y}=\hat{r}\sin{\phi} \sin{\theta}$.
}
\label{fig:TDCEO}
\end{figure}

To solve the polynomial Eqs.~\eqref{eq:omega_TD} and \eqref{eq:W_TD} for $\Omega$ and $W$, we employ a power series expansion in terms of the spin $S$ of the spinning test body. Specifically, we expand $\Omega$ as a series in $S$, reflecting its dependence on the spin. Substituting the series expansion
\beq
\Omega = \Omega_{n}S^n + \mathcal{O}(S^6); \quad W = W_{n}S^n + \mathcal{O}(S^6),
\eeq
into the Eqs.~\eqref{eq:omega_TD} and \eqref{eq:W_TD}, where $\Omega_{n}$ ($n=0,1,2,3,4,5$) are the expansion coefficients to be determined, and $\mathcal{O} (S^6)$ represents the higher terms which are neglected for this approximation. The resulting analytical expressions for the coefficients $\Omega_n(\hat{r},\hat{a},\hat{k})$ for a spinning test body with spin-induced quadrupole under TD SSC, moving on a Kerr background are presented in Tab.~\ref{tab:CEOFrp} and Appendix~\ref{sec:LenghtExp} for the first time and they differ from the numerical findings given in the Appendix of Ref.~\cite{Tim-etal:2023PhRvD:}. Thus, to verify our solutions of the system~\eqref{eq:omega_TD}-\eqref{eq:W_TD} for $\Omega$, we numerically integrate the MPD equations \eqref{eq:spin1}-\eqref{eq:spin3} under TD SSC\footnote{Note that this was not done in Ref.~\cite{Tim-etal:2023PhRvD:}}, employing a fourth-order Gauss implicit Runge-Kutta integrator, and plot in Fig.~\ref{fig:TDCEO} an equatorial circular orbit with the aligned spin for pole-dipole-(spin-induced)quadrupole body. This shows that the frequency $\Omega$ was found correctly.

\begin{table}[ht] 
\centering
   \renewcommand{\arraystretch}{1.4}
  \large
   \begin{tabular}{ |c| c| c| c|}
\hline
$ \hat{\Omega}_n $ & TD SSC & MP SSC  \\   \hline
$\mathcal{O}(\sigma^0)$ & $\frac{1}{\hat{a} \pm \sqrt{\hat{r}^3}}$& $\frac{1}{\hat{a} \pm \sqrt{\hat{r}^3}}$ 
\\ \hline
$\mathcal{O}(\sigma^1)$ &$\frac{3 \left(\pm \hat{a}-\sqrt{\hat{r}}\right)}{2 \sqrt{\hat{r}} \left(\hat{a}\pm \sqrt{\hat{r}^3}\right)^2}$ & $\frac{3 \left(\pm \hat{a}-\sqrt{\hat{r}}\right)}{2 \sqrt{\hat{r}} \left(\hat{a}\pm \sqrt{\hat{r}^3}\right)^2}$  
\\ \hline
$\mathcal{O}(\sigma^2)$ & $\hat{\Omega}_{2}(\hat{r},\hat{a},\hat{k})$ & $\hat{\Omega}_{2}(\hat{r},\hat{a},\hat{k})$
\\ \hline
$\mathcal{O}(\sigma^3)$ & $\hat{\Omega}_{3,\text{TD}}(\hat{r},\hat{a},\hat{k})$ & $\hat{\Omega}_{3,\text{MP}}(\hat{r},\hat{a},\hat{k})$ 
\\ \hline
$\mathcal{O}(\sigma^4)$  &  $\hat{\Omega}_{4,\text{TD}}(\hat{r},\hat{a},\hat{k})$& $\hat{\Omega}_{4,\text{MP}}(\hat{r},\hat{a},\hat{k})$
\\ \hline
$\mathcal{O}(\sigma^5)$  &  $\hat{\Omega}_{5,\text{TD}}(\hat{r},\hat{a},\hat{k})$& $\hat{\Omega}_{5,\text{MP}}(\hat{r},\hat{a},\hat{k})$
\\ \hline
$\mathcal{O}(\sigma^6)$  &  $\hat{\Omega}_{6,\text{TD}}(\hat{r},\hat{a},\hat{k})$& $\hat{\Omega}_{6,\text{MP}}(\hat{r},\hat{a},\hat{k})$
\\ \hline
\end{tabular}
   \caption[caption]{Power series expansion coefficients for orbital frequency $\hat{\hat{\Omega}}_{\pm}$ of spinning test body with spin-induced quadrupole moving in a Kerr background, for TD and MP SSCs. The lengthy expressions of $\hat{\Omega}_{2,3,4}$ for both SSCs can be found in Appendix~\ref{sec:LenghtExp}, while all of  $\hat{\Omega}_{n}$ along with the even lengthier expressions of $\hat{\Omega}_{5,6}$ are given in a Mathematica notebook as supplemental material~\cite{SupMat}.}
   \label{tab:CEOFrp}
\end{table}


\subsubsection{Equatorial circular orbits under MP SSC}\label{ECO_MP} 

Building on the circular equatorial configuration presented in Sec.~\ref{sec:CEOs}, the non-trivial MPD equations under MP SSC read as follows
\begin{align}\label{eq:MP_Str}
& 2 M r^2 \left( 3 S \left( a^2 + r^2 \right) u^{\phi} + a p^{\phi} r^2 - 3 a S u^t - r^2 p^t \right) \nonumber \\ & \times \left( u^t - a u^{\phi} \right) - 3 k M S^2 \left[ 5 a^2 \left( u^t - a u^{\phi} \right)^2 - 3 r^4 (u^{\phi})^2 \right. \nonumber \\ & \left. + r^2 \left( 8 a u^{\phi} \left( a u^{\phi} - u^t \right) + 1 \right) \right] + 2 p^{\phi} r^7 u^{\phi} = 0,
\end{align}
\begin{align}\label{eq:MP_pr}
& r^2 \left( M S \left( u^t - a u^{\phi} \right)^2 + p^{\phi} r^3 u^t \right) - r^5 u^{\phi} \left( p^t + S u^{\phi} \right) \nonumber \\ & + 3 k M S^2 \left[\left( u^t - a u^{\phi} \right) \left( a u^t - \left( a^2 + r^2 \right) u^{\phi} \right)\right] = 0.
\end{align}
We employ the definition of the kinematical mass $m=-p^\alpha u_\alpha$ together with  Eqs.~\eqref{eq:def_omega}, \eqref{eq:utOmega} and \eqref{eq:MP_pr} to derive the analytical expressions for the components $p^t$ and $p^\phi$ as
\begin{widetext}
\bea\non
p^t &=& (2 M m r^5 (a \Omega -1)^2 + S \left(a^2 \Omega  (2 M+r)-2 a M+r^3 \Omega \right) (M (a \Omega -1)   \left(r^2 (a \Omega +3 k S \Omega -1) + 3 a k S (a \Omega -1)\right) - r^5 \Omega^2 ) \\\ &+& m r^6 \left(\Omega^2 \left(a^2+r^2\right)-1\right)) \Big/ \left(r^5 \sqrt{1-\Omega^2 \left(a^2+r^2\right)-\frac{2 M}{r} (a \Omega -1)^2}\, \left[r \left(\Omega ^2 \left(a^2+r^2\right)-1\right)+2 M (a \Omega -1)^2\right]\right), \label{eq:MPpt}\\\non
p^\phi &=& \Big(r^6 \Omega  \left(m \left(\Omega ^2 \left(a^2+r^2\right)-1\right)-S \Omega \right) + 2 M^2 S (a \Omega -1)^2 \left(r^2 (a \Omega +3 k S \Omega -1)+3 a k S (a \Omega -1)\right) \Big. \\\non
&+& \Big. M r (a \Omega -1) \left(r^2 S (a \Omega +3 k S \Omega -1)+3 a k S^2 (a \Omega -1)-2 r^4 \Omega  (m - a m \Omega +S \Omega )\right) \Big)
\\
&\Big/& \left(r^5 \sqrt{1-\Omega^2 \left(a^2+r^2\right)-\frac{2 M}{r} (a \Omega -1)^2}\, \left[r \left(\Omega ^2 \left(a^2+r^2\right)-1\right)+2 M (a \Omega -1)^2\right]\right).\label{eq:MPpphi}
\eea
\end{widetext}
Further substituting the expressions for $p^t$ and $p^\phi$ into Eq.~\eqref{eq:MP_Str} leads to the formulation of the quartic equation for the orbital frequency $\Omega_{\rm MP}$ under MP SSC, given by
\beq\label{eq:MP_omega}
\chi_4\, \Omega_{\rm MP}^4 + \chi_3\,\Omega_{\rm MP}^3 + \chi_2\,\Omega_{\rm MP}^2  + \chi_1\,\Omega_{\rm MP} + \chi_0 =0,
\eeq
where the function $\chi_i$ ($i=0,1,2,3,4$) depends on $r, m ,M, a, S, k$, and can be found in \cite{Tim-etal:2023PhRvD:} along with the solutions of Eq.~\eqref{eq:MP_omega}. For the purpose of comparing the frequencies between TD and MP SSCs as was done in \cite{Timogiannis-etal:2022PhRvD:} for the pole-dipole case, we employ a power series expansion on Eq.~\eqref{eq:MP_omega} in terms of the spin $S$ of the extended body, as described in Sec.~\ref{sec:ECO_TD}, and the resulting expansion coefficients $\Omega_{\rm n |MP}$ under MP SSC are presented in Tab.~\ref{tab:CEOFrp} and Appendix~\ref{sec:LenghtExp}.

\subsection{Comparing $\Omega$ under TD and MP SSC} \label{sec:comparison} 

\begin{figure}
\begin{center}
\includegraphics[width=\hsize]{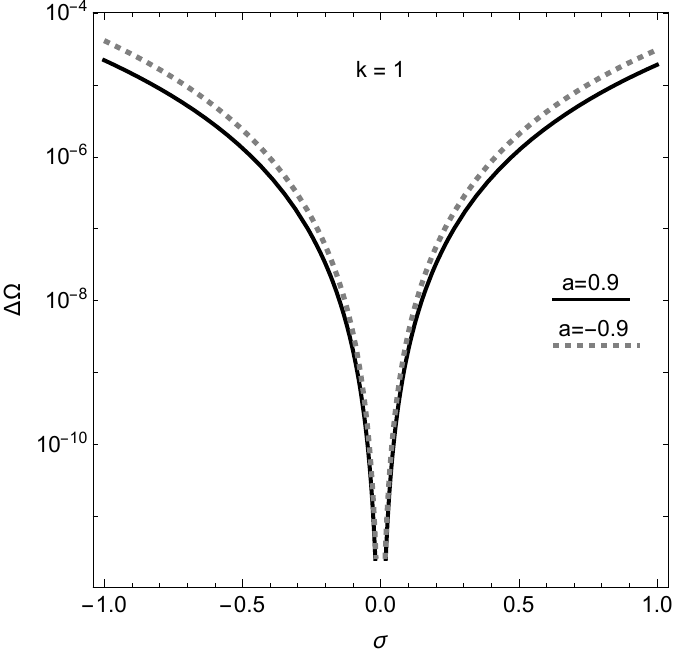}
\end{center}
\caption{The relative difference $\Delta \Omega$ between the orbital frequency under MP SSC $\Omega_{\rm MP}$ and the orbital frequency under TD SSC $\Omega_{\rm TD}$, as a function of spin $\sigma$ for equatorial circular orbit with radius $\hat{r}=10$ of spin-induced extended body orbiting around Kerr BH. To calculate the frequencies, we use the analytical expressions given in Tab.~\ref{tab:CEOFrp} up to $\mathcal{O}(\sigma^6)$ and set $\hat{k}=1$; the solid curve corresponds to Kerr BH spin $\hat{a}=0.9$, while the dashed curve corresponds to $\hat{a}=-0.9$.
}
\label{fig:omega}
\end{figure}

Table~\ref{tab:CEOFrp} provides a comparison of the orbital frequency of the extended body with a pole-dipole-(spin-induced)quadrupole structure moving around a central Kerr BH for prograde and retrograde equatorial circular orbits under TD SSC and MP SSC. The orbital frequencies under both the TD and the MP SSC are equivalent up to the quadratic approximation in spin $\sigma$, i.e., $\Omega_{0, \rm TD} = \Omega_{0, \rm MP}$, $\Omega_{1, \rm TD} = \Omega_{1, \rm MP}$, and $\Omega_{2, \rm TD} = \Omega_{2, \rm MP}$. The lengthy expressions for the power series expansion coefficients for the quadratic and the higher order spin $\sigma$ are provided in the Appendix, where one can see that the TD and the MP SSCs have a stronger level of convergence, being compatible up to $\mathcal{O}(\sigma^2)$-terms. The relative difference $\Delta \Omega$ between the orbital frequency under MP SSC $\Omega_{\rm MP}$ and the orbital frequency under TD SSC $\Omega_{\rm TD}$ is given by
\beq
\Delta \Omega = \left| 1 - \frac{\Omega_{\rm MP}}{\Omega_{\rm TD}} \right|,
\eeq
and presented in Fig.~\ref{fig:omega}, as a function of test extended body spin $\sigma$, for $\hat{k}=1$, $\hat{a}=\pm0.9$ and $\hat{r}=10$. The frequencies are calculated with up to $\mathcal{O}(\sigma^6)$ approximation. The orbital frequencies presented in Tab.~\ref{tab:CEOFrp} do not incorporate the centroid's correction. In the following section, we examine how the orbital frequencies $\Omega$ under TD and MP SSCs are affected when centroid corrections are applied.

\section{Centroid's comparison} \label{sec:SSCcomp}

When a centroid shifts from $z^\alpha$ to $ \tilde{z}^{\alpha} = z^\alpha + \delta z^\alpha$ due a SSC change, the spin tensor transforms as
\beq\label{eq:tild_spin_tensor}
   \tilde{S}^{\alpha \beta} = S^{\alpha \beta} + p^\alpha \delta z^\beta - p^\beta \delta z^\alpha,
\eeq
while the contravariant components of the four-momentum $p^\alpha$ remain unchanged by this shift in the worldline, i.e., $\tilde{p}^\alpha =p^\alpha$, with the covariant components given by $\tilde{p}_\alpha = \tilde{g}_{\alpha \beta}\, p^\beta$, where $\tilde{g}_{\alpha \beta}$ represents the metric evaluated at the shifted radial coordinate \(\tilde{r} = r + \delta r\). The shift $\delta z^\alpha$ is expressed as \cite{Kyrian-Semerak:2007:MNRAS:}
\beq\label{eq:z_mu}
\delta z^\alpha = \frac{\tilde{p}_\beta S^{\beta \alpha}}{\tilde{\mu}^2},
\eeq
where \(\tilde{\mu}^2 = -\tilde{g}_{\nu \sigma} p^\nu p^\sigma\) is the dynamical rest mass. During a centroid shift, the quadrupole tensor also undergoes a transformation \cite{Vines:2016PhRvD} that is not addressed in our study.

In our work, the quantities with a tilde over refer to those calculated at the TD frame, while the quantities without a tilde refer to those calculated at the MP centroid. We shift from the MP SSC frame to the TD SSC frame, since the TD centroid is uniquely defined, while the MP centroid is not. If we had opted to shift from TD to MP SSC, we would have to address the question of whether the MP centroid undergoes helical motion \cite{Filipe-etal:2018:PhRvD:}. By opting for the non-helical MP to TD transition, we do not have to address the aforementioned issue.     

\begin{table}[ht] 
\centering
   \renewcommand{\arraystretch}{1.4}
  \large
   \begin{tabular}{ |c| c| c| }
\hline

$ \hat{\Omega}_n $ & TD SSC & MP SSC  \\   \hline
$\mathcal{O}(\sigma^0)$ & $\frac{1}{\hat{a} \pm \sqrt{\hat{r}^3}}$& $\frac{1}{\hat{a} \pm \sqrt{\hat{r}^3}}$ 
\\ \hline
$\mathcal{O}(\sigma^1)$ &$\frac{3 \left(\pm \hat{a}-\sqrt{\hat{r}}\right)}{2 \sqrt{\hat{r}} \left(\hat{a}\pm \sqrt{\hat{r}^3}\right)^2}$ & $\frac{3 \left(\pm \hat{a}-\sqrt{\hat{r}}\right)}{2 \sqrt{\hat{r}} \left(\hat{a}\pm \sqrt{\hat{r}^3}\right)^2}$  
\\ \hline
$\mathcal{O}(\sigma^2)$ & $\hat{\Omega}_{2}(\hat{r},\hat{a},\hat{k})$ & $\hat{\Omega}_{2}(\hat{r},\hat{a},\hat{k})$
\\ \hline
$\mathcal{O}(\sigma^3)$ & $\hat{\Omega}_{3,\text{MP}}(\hat{r},\hat{a},\hat{k})$ & $\hat{\Omega}_{3,\text{MP}}(\hat{r},\hat{a},\hat{k})$ 
\\ \hline
$\mathcal{O}(\sigma^4)$  &  $\hat{\Omega}_{4,\text{TD}}^{'}(\hat{r},\hat{a},\hat{k})$& $\hat{\Omega}_{4,\text{MP}}(\hat{r},\hat{a},\hat{k})$
\\ \hline
$\mathcal{O}(\sigma^5)$  &  $\hat{\Omega}_{5,\text{TD}}^{'}(\hat{r},\hat{a},\hat{k})$& $\hat{\Omega}_{5,\text{MP}}(\hat{r},\hat{a},\hat{k})$
\\ \hline
$\mathcal{O}(\sigma^6)$  &  $\hat{\Omega}_{6,\text{TD}}^{'}(\hat{r},\hat{a},\hat{k})$& $\hat{\Omega}_{6,\text{MP}}(\hat{r},\hat{a},\hat{k})$
\\ \hline
\end{tabular}
   \caption[caption]{
Power series expansion coefficients for orbital frequency $\hat{\hat{\Omega}}_{\pm}$ of spinning test body with spin-induced quadrupole moving in a Kerr background, after applying radial linear corrections to the position of the center of mass defined by MP SSC when we change the SSC to the TD one. The lengthy expressions of $\hat{\Omega}_{2,3}$ for both SSCs along with $\hat{\Omega}_{4,\text{TD}}^{'}$ can be found in Appendix~\ref{sec:LenghtExp}, while all the $\hat{\Omega}_{n,\text{TD}}^{'}$ including the even lengthier $\hat{\Omega}_{5,\text{TD}}^{'},\hat{\Omega}_{6,\text{TD}}^{'}$ can be found in \cite{SupMat}.
}
   \label{tab:radial_shifts}
\end{table}


\begin{table}[h] 
\centering
   \renewcommand{\arraystretch}{1.4}
  \large
   \begin{tabular}{ |c| c| c| }
\hline

$ \hat{\Omega}_n $ & zero-order $\delta r$ & first-order $\delta r$  \\   \hline

$\mathcal{O}(\sigma^4)$ & $ \hat{\Omega}_{4,\text{TD}}^{'} = \hat{\Omega}_{4,\text{MP}}$& $\hat{\Omega}_{4,\text{TD}}^{'} = \hat{\Omega}_{4,\text{MP}}$ 
\\ \hline
$\mathcal{O}(\sigma^5)$ &$ \hat{\Omega}_{5,\text{TD}}^{'} = \hat{\Omega}_{5,\text{MP}}$ & $\hat{\Omega}_{5,\text{TD}}^{'} \neq \hat{\Omega}_{5,\text{MP}}$  
\\ \hline
$\mathcal{O}(\sigma^6)$ & $ \hat{\Omega}_{6,\text{TD}}^{'} \neq \hat{\Omega}_{6,\text{MP}} $ & $ \hat{\Omega}_{6,\text{TD}}^{'} \neq \hat{\Omega}_{6,\text{MP}} $
\\ \hline
\end{tabular}
   \caption[caption]{Comparison of the frequency components for $\hat{k}=1$ under zero-order estimate in $\delta r$ and zero-order estimate in $\delta r$, when only the linear radial shift is taken into account.
}
   \label{tab:Radial_k1}
\end{table}


\subsection{Radial linear corrections} \label{sec:RadCor}

To address the divergence in the power series of orbital frequencies observed between the SSCs, as presented in Table \ref{tab:CEOFrp}, we apply a linear correction to the position 
\beq \label{eq:RadShift}
\tilde{r} = r + \delta r,
\eeq
where $\delta r$ is given by Eq.~\eqref{eq:z_mu}. At this point, we keep the spin measure of the test body unchanged under the shift in SSC, i.e., $\tilde{\sigma}=\sigma$. The computation of the centroid's radial shift may seem complex, as the right-hand side of Eq.~\eqref{eq:z_mu} depends on $\delta r$ through the four-momentum $\tilde{p}_\alpha$ and the dynamical rest mass $\tilde{\mu}$. By expanding Eq.~\eqref{eq:z_mu} within the linear approximation in terms of $\delta r$, one obtains \cite{Ias:Gera:The:2021:PhRvD:,Timogiannis-etal:2022PhRvD:} 
\bea\non
\delta r & \approx & \frac{p_t S^{tr} + p_\phi S^{\phi r}}{\mu^2} + \delta r \left[\frac{g_{t t, r} p^t S^{tr} + g_{\phi \phi, r} p^\phi S^{\phi r}}{\mu^2} \right.\\\non &+& \left. \frac{g_{t \phi, r} (S^{tr} p^\phi + S^{\phi r} p^t)}{\mu^2} +\frac{(p_t S^{tr} + p_\phi S^{\phi r})}{\mu^4} \right. \\\non &\times& \left. \left\{ g_{tt,r} (p^t)^2 + 2g_{t\phi,r} p^t p^\phi + g_{\phi \phi,r} (p^\phi)^2 \right \} \right] \\&+& \mathcal{O}(\delta r^2). \label{eq:delta_r} 
\eea
To obtain a fundamental approximation of the radial shift, we neglect the $\delta r$-term on the right-hand side of Eq.~\eqref{eq:delta_r}, providing a ``zero-order'' estimate for the correction to the centroid's position. When we take the $\delta r$-term on the right-hand side into account, we call it a ``first-order'' estimate. We then substitute the components of four-momentum $p_\alpha$ and spin-tensor $S^{\alpha \beta}$ into Eq.~\eqref{eq:delta_r}, computed using a system of Eqs.~\eqref{eq:MP_Str}-\eqref{eq:MP_pr}, with $p_t=g_{\alpha t}\, p^\alpha$, and $p_\phi = g_{\alpha \phi}\, p^\phi$, while the $S^{tr}$ and $S^{r \phi}$ are computed using Eqs.~\eqref{eq:ST-1} and \eqref{eq:ST-2}, with $V^\alpha = u^\alpha$. The relation $u^\phi = \Omega\, u^t$ and Eq.~\eqref{eq:utOmega} are taken into account, where $\Omega$ corresponds to the MP SSC. The zero-order estimate is given by
\begin{widetext}
\beq
\delta r = -\frac{\sigma ^2 \left(\hat{a}^2+\left(\hat{r}-2\right) \hat{r}\right) \left(\hat{r}^2 \left(\hat{a} \hat{\Omega} -1\right) \left(\hat{a} \hat{\Omega} +3 \hat{k} \sigma  \hat{\Omega} -1\right)+3 \hat{a} \hat{k} \sigma  \left(\hat{a} \hat{\Omega} -1\right)^2+\left(-\hat{r}^5\right) \hat{\Omega} ^2\right)}{\hat{r}^5 \left(\hat{\Omega} ^2 \left(\hat{a}^2 \left(\hat{r}+2\right)+\hat{r}^3\right)-4 \hat{a} \hat{\Omega} -\hat{r}+2\right)},
\eeq
\end{widetext}
while the first-order is given in the supplemental material~\cite{SupMat}. 

After computing $\delta r$, we substitute $\tilde{r} = r + \delta r$ into the second column (i.e. to $\hat{\Omega}_{\rm n,TD}$) of Tab.~\ref{tab:CEOFrp}, to address the discrepancies observed in the orbital frequency under TD and MP SSCs. The results obtained after shifting the centroids are presented in Tab.~\ref{tab:radial_shifts}. Introducing the centroid shift removes the discrepancy between the orbital frequencies under the TD and MP SSCs up to the $\mathcal{O}(\sigma^3)$-term. For the $\hat{\Omega}_4$ terms, we notice that for both $\delta r$ estimates (zero \& first order), $\hat{\Omega}_n$ are same upto $O(S^3)$, but changes after $O(S^3)$. Namely by applying the zero-order estimate for $\delta r$, we get 
\begin{widetext}
\beq\label{eq:dif-I}
\hat{\Omega}^{'}_{4,\rm TD}-\hat{\Omega}_{4,\rm MP} = \frac{9 (\hat{k}-1)^2 \left(\hat{a} \mp \sqrt{\hat{r}}\right) \left(\pm 5 \hat{a}^3-3 \hat{a}^2 \sqrt{\hat{r}} \pm \hat{a} (4 \hat{r}-9) \hat{r} + (5-2 \hat{r}) \hat{r}^{3/2}\right)}{2 \hat{r}^6 \left(\pm 2 \hat{a} + (\hat{r}-3) \sqrt{\hat{r}}\right) \left(\hat{a} \pm \hat{r}^{3/2}\right)^2}.
\eeq
\bea \label{eq:dif-II} \non
\hat{\Omega}^{'}_{5,\rm TD}-\hat{\Omega}_{5,\rm MP} &=& \frac{9(\hat{k}-1)}{4 \hat{r}^8 \left(\pm 2 \hat{a}+(\hat{r}-3) \sqrt{\hat{r}}\right)^2 \left(\hat{a} \pm \hat{r}^{3/2}\right)^3}
[ 120 \hat{a}^7 (1-2 \hat{k}) \pm \hat{a}^6 \sqrt{\hat{r}} (\hat{k} (753-285 \hat{r})+105 \hat{r}-353) \\\non &+& 
\hat{a}^5 \hat{r} \left(15 (1-5 \hat{k}) \hat{r}^2+459 \hat{k} \hat{r}-268 \hat{k}-69 \hat{r}+78\right) \pm 2 \hat{a}^4 \hat{r}^{3/2} (\hat{k} ((395-87 \hat{r}) \hat{r}-612)+\hat{r} (52 \hat{r}-269) \\\non &+& 301)+2 \hat{a}^3 \hat{r}^2 (\hat{k} (649-3 \hat{r} (4 \hat{r} (3 \hat{r}-31)+311))+\hat{r} (3 \hat{r} (2 \hat{r}-37)+365)-272) \pm \hat{a}^2 \hat{r}^{5/2} (7 \hat{k} (\hat{r} (8 \hat{r}^2 
\\\non &-& 60 \hat{r}+109 )-29 )+\hat{r} (4 \hat{r} (4 \hat{r}-17)+45)+31 )+\hat{a} \hat{r}^3 (\hat{k} \left(\hat{r} \left(\hat{r} \left(-8 \hat{r}^2+84 \hat{r}-281\right)+339\right)-114\right) \\ &+& \hat{r} ((295-52 \hat{r}) \hat{r}-429)+66 )+2 (\hat{r}-3) \hat{r}^{9/2} (\hat{k} (\hat{r}-2) (3 \hat{r}-17)+12 \hat{r}-26)]. 
\eea
After applying the first-order estimate, we have
\beq\label{eq:dif-III}
\hat{\Omega}^{'}_{4,\rm TD}-\hat{\Omega}_{4,\rm MP} = 
\frac{9 (\hat{k}-1) \left(\hat{a}-\sqrt{\hat{r}}\right) \left(5 \hat{a}^3 (\hat{k}-1)+\hat{a}^2 (5-3 \hat{k}) \sqrt{\hat{r}}+\hat{a} (\hat{k}-1) (4 \hat{r}-9) \hat{r}+\hat{r}^{3/2} (\hat{k} (5-2 \hat{r})+4 \hat{r}-9)\right)}{2 \hat{r}^6 \left(2 \hat{a}+(\hat{r}-3) \sqrt{\hat{r}}\right) \left(\hat{a}+\hat{r}^{3/2}\right)^2}.
\eeq
\bea \label{eq:dif-iv}  \non
\hat{\Omega}^{'}_{5,\rm TD}-\hat{\Omega}_{5,\rm MP} &=& \frac{9}{4 \hat{r}^8 \left(2 \hat{a}+(\hat{r}-3) \sqrt{\hat{r}}\right)^2 \left(\hat{a}+\hat{r}^{3/2}\right)^3}[-120 \hat{a}^7 (\hat{k}-1) (2 \hat{k}-1) \hat{a}^6 \sqrt{\hat{r}} (753 \hat{k}^2-15 (\hat{k}-1) (19 \hat{k}-7) \hat{r} \\\non &-& 1134 \hat{k}+361)+\hat{a}^5 \hat{r} \left(\hat{k}^2 \left(-75 \hat{r}^2+459 \hat{r}-268\right)+6 \hat{k} (\hat{r} (15 \hat{r}-89)+64)-15 (\hat{r}-3) \hat{r}-62\right) \\\non &-& 2 \hat{a}^4 \hat{r}^{3/2} \left(\hat{k}^2 (\hat{r} (87 \hat{r}-395)+612)+\hat{k} ((713-144 \hat{r}) \hat{r}-959)+\hat{r} (62 \hat{r}-339)+345\right)-2 \hat{a}^3 \hat{r}^2\\\non &\times& \left(\hat{k}^2 (3 \hat{r} (4 \hat{r} (3 \hat{r}-31)+311)-649)+\hat{k} (\hat{r} (7 (73-6 \hat{r}) \hat{r}-1404)+983)+\hat{r} (\hat{r} (6 \hat{r}-127)+429)-280\right) \\\non &+& \hat{a}^2 \hat{r}^{5/2} (7 \hat{k}^2 \left(\hat{r} \left(8 \hat{r}^2-60 \hat{r}+109\right)-29\right)+\hat{k} (162-2 \hat{r} (8 \hat{r} (2 \hat{r}-19)+313))-3 \hat{r} (4 \hat{r} (3 \hat{r}-17)+99) \\\non &+& 113) + \hat{a} \hat{r}^3 (\hat{k}^2 \left(\hat{r} \left(\hat{r} \left(-8 \hat{r}^2+84 \hat{r}-281\right)+339\right)-114\right) + 2 \hat{k} (\hat{r} (\hat{r} (\hat{r} (4 \hat{r}-93)+419)-570)+138) \\\non &+& \hat{r} (\hat{r} (108 \hat{r}-599)+861)-162)+2 \hat{r}^{9/2} (\hat{k}^2 (\hat{r}-3) (\hat{r}-2) (3 \hat{r}-17)+\hat{k} (\hat{r} ((67-4 \hat{r}) \hat{r}-249)+264)\\ &-& 2 (5 \hat{r} (3 \hat{r}-14)+81))]. 
\eea
\end{widetext}
For $\hat{k}=1$, Eqs.~\eqref{eq:dif-I}-\eqref{eq:dif-III} vanish, but not Eq.~\eqref{eq:dif-iv}. In both the zero-order estimate and the first-order estimate cases of $\delta r$ when $\hat{k}=1$, we have $\hat{\Omega}^{'}_{6,\rm TD}\neq\hat{\Omega}_{6,\rm MP}$. The interested reader is referred to the supplemental material~\cite{SupMat} to check this latter lengthy result. Hence, for $\hat{k}=1$, the convergence between the two SSCs increases to $\mathcal{O}(\sigma^5)$-term for the zero-order estimate of $\delta r$, but for the first-order estimate of $\delta r$, it remains at $\mathcal{O}(\sigma^4)$-term (see Tab.~\ref{tab:Radial_k1}). If we take into account that the generalized Carter constant in the pole-dipole-(spin-induced)quadrupole approximation found in \cite{Compere:2023ScPP} is claimed to exist just when $\hat{k}=1$, then one can speculate that BHs possess some extra property than the other compact objects. We shall take into account that the zero-order estimate allows the frequencies between the TD and MP SSCs to converge more and limit our detailed presentation of formulas to the zero-order estimate in the next section, in which the spin correction will be incorporated.

\begin{table}[t] 
\centering
   \renewcommand{\arraystretch}{1.4}
  \large
   \begin{tabular}{ |c| c| c| }
\hline

$ \hat{\Omega}_n $ & TD SSC & MP SSC  \\   \hline

$\mathcal{O}(\sigma^0)$ & $\frac{1}{\hat{a} \pm \sqrt{\hat{r}^3}}$& $\frac{1}{\hat{a} \pm \sqrt{\hat{r}^3}}$ 
\\ \hline
$\mathcal{O}(\sigma^1)$ &$\frac{3 \left(\pm \hat{a}-\sqrt{\hat{r}}\right)}{2 \sqrt{\hat{r}} \left(\hat{a}\pm \sqrt{\hat{r}^3}\right)^2}$ & $\frac{3 \left(\pm \hat{a}-\sqrt{\hat{r}}\right)}{2 \sqrt{\hat{r}} \left(\hat{a}\pm \sqrt{\hat{r}^3}\right)^2}$  
\\ \hline
$\mathcal{O}(\sigma^2)$ & $\hat{\Omega}_{2}(\hat{r},\hat{a},\hat{k})$ & $\hat{\Omega}_{2}(\hat{r},\hat{a},\hat{k})$
\\ \hline
$\mathcal{O}(\sigma^3)$ & $\hat{\Omega}_{3,\text{MP}}(\hat{r},\hat{a},\hat{k})$ & $\hat{\Omega}_{3,\text{MP}}(\hat{r},\hat{a},\hat{k})$ 
\\ \hline
$\mathcal{O}(\sigma^4)$  &  $\hat{\Omega}_{4,\text{TD}}^{''}(\hat{r},\hat{a},\hat{k})$& $\hat{\Omega}_{4,\text{MP}}(\hat{r},\hat{a},\hat{k})$
\\ \hline
$\mathcal{O}(\sigma^5)$  &  $\hat{\Omega}_{5,\text{TD}}^{''}(\hat{r},\hat{a},\hat{k})$& $\hat{\Omega}_{5,\text{MP}}(\hat{r},\hat{a},\hat{k})$
\\ \hline
\end{tabular}
   \caption[caption]{
Power series expansion coefficients for orbital frequency $\hat{\hat{\Omega}}_{\pm}$ of spinning test body with spin-induced quadrupole moving in a Kerr background, after applying spin measure corrections for the TD and MP SSCs. The lengthy expressions of $\hat{\Omega}_{n,\text{TD}}^{''}$ can be found in \cite{SupMat}.
}
   \label{tab:spin_shifts}
\end{table}


\begin{table}[t] 
\centering
   \renewcommand{\arraystretch}{1.4}
  \large
   \begin{tabular}{ |c| c| c| }
\hline

$ \hat{\Omega}_n $ & zero-order $\delta r$ & first-order $\delta r$  \\   \hline

$\mathcal{O}(\sigma^4)$ & $ \hat{\Omega}_{4,\text{TD}}^{''} = \hat{\Omega}_{4,\text{MP}}$& $\hat{\Omega}_{4,\text{TD}}^{'} = \hat{\Omega}_{4,\text{MP}}$ 
\\ \hline
$\mathcal{O}(\sigma^5)$ &$ \hat{\Omega}_{5,\text{TD}}^{''} \neq \hat{\Omega}_{5,\text{MP}}$ & $\hat{\Omega}_{5,\text{TD}}^{'} \neq \hat{\Omega}_{5,\text{MP}}$  
\\ \hline
\end{tabular}
   \caption[caption]{Comparison of the frequency components for $\hat{k}=1$ under zero-order estimate in $\delta r$ and zero-order estimate in $\delta r$, when both the linear radial and the shift in spin correction are performed.
}
   \label{tab:RadialSpin_k1}
\end{table}


\subsection{Spin shift corrections} \label{sec:SpinCor}

A comprehensive analysis of the behavior of the centroid of a spinning test body must account for the strongly SSC-dependent nature of the spin measure. In other words, the transition between different SSCs shifts the representative worldline with respect to which the moments are evaluated, thereby affecting the spin measure itself. In this section, we assume that $\tilde{\sigma}\neq\sigma$, where $\tilde{\sigma} = \frac{S}{\tilde{\mu} M}$, $\sigma = \frac{S}{m M}$, and compute the corresponding power series expansions for the orbital frequencies $\hat{\Omega}_{\pm}$ under TD and MP SSCs. This is accomplished by combining Eqs.~\eqref{eq:spin_magnitude} and \eqref{eq:tild_spin_tensor} for a spinning test body moving in circular equatorial orbits around a Kerr BH. Recall that for such orbits, the only non-zero components of the spin-tensor are $S^{t r} = - S^{r t}$, and $S^{r \phi}=-S^{\phi r}$. The expansion procedure described here yields \cite{Ias:Gera:The:2021:PhRvD:,Timogiannis-etal:2022PhRvD:}
\bea\non
\tilde{S}^2 &=& S^2 + \delta r \left[ g_{rr} \left( g_{\phi\phi, r} \left(S^{r\phi} \right)^2 - 2 g_{t \phi,r} S^{tr} S^{r \phi}  \right.\right.\\\non &+& \left.\left. g_{tt, r} \left(S^{tr} \right)^2 \right) + 2 g_{rr} \left(p_t S^{tr} - p_{\phi} S^{r \phi} \right) \right. \\\label{tild_spin} &+& \left. \frac{S^2}{g_{rr}} g_{rr,r} \right] + \mathcal{O} (\delta r^2).
\eea
In the next step of the derivation, we normalize all the quantities appearing in Eq.~\eqref{tild_spin} to make them dimensionless.

In order to obtain a relation $\tilde{\sigma}=f(\sigma)$, for a transition from the MP to the TD centroid, we divide both sides of Eq.~\eqref{tild_spin} by $\tilde{\mu}^2 M^2$. Using the definition of dynamical rest mass $\tilde{\mu}^2 = - \tilde{g}_{\alpha \beta}\, p^\alpha p^\beta$, we obtain the inverse square of the dynamical rest mass in the linear approximation with respect to radial shift \( \delta r \), given by 
\bea\non
\frac{1}{\tilde{\mu}^2} &=& \frac{1}{\mu^2} + \frac{\delta r}{\mu^4} \left[g_{tt,r}\, \left(p^t\right)^2 + 2 g_{t \phi, r}\, p^t p^\phi \right.\\\label{tild_mu} &+& \left. g_{\phi \phi ,r}\, \left(p^\phi\right)^2 \right] + \mathcal{O}(\delta r^2).
\eea
Recall that the dimensionless spin under the MP SSC is given by $\sigma = \frac{S}{m M}$, while the dimensionless spin under TD SSC is $\tilde{\sigma} = \frac{S}{\tilde{\mu} M}$. Since $\sigma$ and $\tilde{\sigma}$ are not defined identically, it is necessary to relate the dynamical rest mass $\tilde{\mu}$ to the kinematical mass $m$, both measured in the MP reference frame. This relation reads 
\beq\label{mass_relation_1}
\mu^2 = m^2 + \frac{1}{S^2} \left[S^{\alpha \beta} S_{\beta \gamma}\, p^{\gamma} p_{\alpha} - \left(p_{\alpha} S^{\alpha}\right)^2 \right].
\eeq
Substituting the components of spin-tensor and four-momentum corresponding to circular equatorial orbits into Eq.~\eqref{mass_relation_1}, we obtain \cite{Timogiannis-etal:2022PhRvD:} 
\beq\label{mass_relation_2}
\mu^2 = \frac{1}{S^2} \left[m^2 S^2 - g_{rr} \left(p_{\phi} S^{r \phi} - p_t S^{tr} \right)^2 \right].
\eeq
By inserting Eq.~\eqref{mass_relation_2} into Eq.~\eqref{tild_mu}, we express the dynamical rest mass $\tilde{\mu}$ measured in the TD frame as a function of the kinematical mass $m$ measured in the MP frame. The resulting relation is given by 
\bea\non
\frac{1}{\tilde{\mu}^2} &=& \frac{S^2}{m^2 S^2 - g_{rr} \left(p_{\phi} S^{r \phi} - p_t S^{tr} \right)^2} \\\non &+& \frac{\delta r\, S^4}{\left(m^2 S^2 - g_{rr} \left(p_{\phi} S^{r \phi} - p_t S^{tr} \right)^2 \right)^4} \left[g_{tt,r}\, \left(p^t\right)^2 \right.\\\ &+& \left. 2 g_{t \phi, r}\, p^t p^\phi + g_{\phi \phi ,r}\, \left(p^\phi\right)^2 \right] + \mathcal{O}(\delta r^2).
\eea
Consequently, the desired relation between $\tilde{\sigma}$ and $\sigma$, expressed as a power series expansion takes the form 
\begin{widetext}

\bea\non
\tilde{\sigma}^2 &=& \frac{\sigma^4}{ \sigma^2 - \frac{g_{rr}}{m^2} \left(p_t \sigma^{tr} - p_\phi \sigma^{r \phi} \right)^2} + \frac{\delta r \, \sigma^6 \left[g_{tt,r}\, \left(\frac{p^t}{m} \right)^2 + 2 g_{t \phi, r}\, \frac{p^t p^\phi}{m^2} + g_{\phi \phi ,r}\, \left(\frac{p^\phi}{m} \right)^2 \right]}{\left(\sigma^2 - \frac{g_{rr}}{m^2} \left(p_t\, \sigma^{tr} - p_\phi\, \sigma^{r \phi} \right)^2 \right)^2 } \\\label{tild_sigma} &+& \frac{ \delta r\, \sigma^2 \left[ g_{rr} \left\{ g_{\phi\phi, r} (\sigma^{r \phi})^2 +2 g_{t \phi, r} \sigma^{tr} \sigma^{r \phi}  + g_{tt, r} (\sigma^{tr})^2 + \frac{2}{m M} \left(p_t \sigma^{tr} - p_{\phi} \sigma^{r \phi} \right) \right\}  + \frac{\sigma^2}{g_{rr}} g_{rr,r} \right] }{\sigma^2 - \frac{g_{rr}}{m^2} \left(p_t \sigma^{tr} - p_\phi \sigma^{r \phi} \right)^2} + \mathcal{O}(\delta r^2), 
\eea
\end{widetext}
where the normalized spin-tensor $\sigma^{\alpha \beta}=\frac{S^{\alpha \beta}}{m M }$ has been used. By substituting the expression for $\delta r$ and four-momentum into Eq.~\eqref{tild_sigma}, and applying power series expansion with respect to $\sigma$, we obtain after applying zero-order estimate of $\delta r$:
\begin{widetext}
\bea\non
\tilde{\sigma} &=& \sigma + \frac{3 \sigma ^4 \left(\hat{k}-1\right) \left(\sqrt{\hat{r}} \mp \hat{a}\right)}{\hat{r}^8 \left(\pm 2 \hat{a}+\left(\hat{r}-3\right) \sqrt{\hat{r}}\right)^2} 
\left[- 2 \hat{a} \left(\hat{r}-2\right)^2 \hat{r}^{5/2} \mp \hat{a}^2 \left(\hat{r}^3-3 \hat{r}^2+4 \hat{r}-16\right) \hat{r} - 2 \hat{a}^3 \left(\hat{r}^2-2 \hat{r}+8\right) \sqrt{\hat{r}} \pm 4 \hat{a}^4 \right. \\\non  &\mp& \left. \left(\hat{r}^2-5 \hat{r}+6\right) \hat{r}^4 \right] 
+ \frac{3 \sigma^5}{2 \hat{r}^{10} \left(\pm 2 \hat{a}+\left(\hat{r}-3\right) \sqrt{\hat{r}}\right)^3} 
[\pm 40 \hat{a}^7 \hat{k} +  \left(6 -\hat{k} \left(\hat{r}-5\right) \left(\hat{r}-3\right)+9 \hat{k}^2-5 \hat{r}\right) \left(\hat{r}-3\right) \left(\hat{r}-2\right) \\\non &\times& \hat{r}^{11/2}  \mp 2 \hat{a}^5 \hat{r} \left(\left(3-13 \hat{k}\right) \hat{r}^3+\left(9 \hat{k}^2+57 \hat{k}-14\right) \hat{r}^2-122 \hat{k} \hat{r}+394 \hat{k}-62\right) + \hat{a}^2 \hat{r}^{5/2} \left(-6 \hat{k} \hat{r}^5+3 \left(3 \hat{k} \left(\hat{k}+7\right) \right. \right. \\\non &-& \left. \left. 2\right) \hat{r}^4-2 \left(4 \hat{k} \left(9 \hat{k}+23\right)-5\right) \hat{r}^3+\left(\hat{k} \left(99 \hat{k}+203\right)+64\right) \hat{r}^2-12 \left(9 \hat{k}+7\right) \hat{r}+48 \left(\hat{k}-1\right)\right) + 4 \hat{a}^6 \sqrt{\hat{r}} \left(\hat{k} \left(\left(15 \right. \right. \right. \\\non &-& \left. \left. \left. 8 \hat{r}\right) \hat{r}-75\right)+3 \hat{r}^2+8\right) \mp \hat{a} \hat{r}^4 \left(\hat{r}-2\right)  \left(18 \hat{k}^2 \left(\hat{r}-4\right) \hat{r}+\hat{k} \left(\hat{r} \left(3 \left(\hat{r}-8\right) \hat{r}+77\right)-48\right)-\left(5 \hat{r}-16\right) \left(\left(\hat{r}-2\right) \hat{r}+3\right)\right) \\\non &\mp& \hat{r}^2  \hat{a}^3 \left(36 \hat{k}^2 \hat{r}^2+\hat{k} \left(\hat{r} \left(\hat{r} \left(\left(190-29 \hat{r}\right) \hat{r}-405\right)+188\right)+272\right)+\left(\hat{r} \left(11 \left(\hat{r}-6\right) \hat{r}+95\right)-60\right) \hat{r}+16\right) 
+\hat{a}^4 \hat{r}^{3/2} \\ &\times& \left(9 \hat{k}^2 \left(\hat{r}-7\right) \hat{r}^2+\hat{k} \left(\hat{r} \left(\hat{r} \left(31-5 \left(\hat{r}-2\right) \hat{r}\right)+184\right)-844\right)+\left(\hat{r} \left(11 \hat{r}-37\right)+48\right) \hat{r}+124\right)
] + \mathcal{O}(\sigma^6),\label{eq:spin_correction} 
\eea
\end{widetext}
which reduces to the pole-dipole case for vanishing $k$ \cite{Timogiannis-etal:2022PhRvD:}.\footnote{The supplementary material~\cite{SupMat} provides the respective  Eq.~\eqref{eq:spin_correction} when the first-order estimate is applied.} By inserting Eq.~\eqref{eq:spin_correction} into the second column of Tab.~\ref{tab:radial_shifts}, and applying power series expansion with respect to $\sigma$, we derive the coefficients of power series expansion of orbital frequency presented in Tab.~\ref{tab:spin_shifts}. It is obvious from Eq.~\eqref{eq:spin_correction} and Tab.~\ref{tab:spin_shifts} that the coefficients of the orbital frequency satisfying the inequality $\hat{\Omega}_{n} \leq \mathcal{O}(\sigma^3)$ are not influenced by the alteration of the spin. By inserting the function $\tilde{\sigma}=f(\sigma)$ into the term $\hat{\Omega}_{4,\text{TD}}^{'}(\hat{r},\hat{a},\hat{k})$, we obtain the quantity $\hat{\Omega}_{4,\text{TD}}^{''}(\hat{r},\hat{a},\hat{k})$, which differs from the term $\hat{\Omega}_{4,\text{MP}}(\hat{r},\hat{a},\hat{k})$. This observation clearly indicates that further improvement in the convergence of the frequency power series is not possible. Also notice that for $\hat{k}=1$, Eq.~\eqref{eq:spin_correction} reduces to $\tilde{\sigma}=\sigma+\mathcal{O}(\sigma^5)$. When we compare the frequencies for $\hat{k}=1$, the convergence again rises to $\mathcal{O}(\sigma^4)$ as shown in Tab.\ref{tab:RadialSpin_k1} for both the zero-order and the first-order estimate of $\delta r$. It is counterintuitive that the zero-order estimate of $\delta r$ provided even higher convergence for $\hat{k}=1$, while the more complete shifts due to the change of the SSC failed to do so. This fact along with that the spin correction did not improve the  convergence in the pole-dipole case~\cite{Timogiannis-etal:2022PhRvD:} discourages us from trying to incorporate the quadrupole correction into our study.

\section{Innermost stable circular orbits}\label{sec:ISCOs} 

The ISCOs represent the boundary of stability for circular equatorial orbits of a particle around BHs. Specifically, orbits with $r<r_{\rm ISCO}$ are unstable, whereas those with $r>r_{\rm ISCO}$ are stable. Various methods have been developed to determine the radius of ISCOs, spanning from the analytical work of Bardeen et al. \cite{Bardeen-etal:1972ApJ:} on geodesic orbits to more recent advancements, such as those presented in \cite{Stein-Niels:2020PhRvD:}. For an extended body described by the MPDs in the pole-dipole approximation the attempts start with \cite{Rasband1973PhRv} and \cite{Tod-etal:1976:NCimB:} to more recent ones \cite{Steinhoff-Dirk:2012:PhRvD:,Jefremov2015PhRvD,Harms2016PhRvD,Timogiannis-etal:2022PhRvD:}, while for the pole-dipole-(spin-induced)quadrupole case we have just Refs.~\cite{Steinhoff-Dirk:2012:PhRvD:,Bini-Andrea:2015PhRvD:} up to certain appropriation in particle spin $S$ and BH spin parameter $S$.

\subsubsection{ISCOs under TD SSC}\label{sec:ISCO_TD}  

To find the ISCOs under TD SSC, we replace the reference four-vector $V^\alpha=p^\alpha/\mu$ in Eqs.~\eqref{eq:energy} and \eqref{eq:Jz}, the expressions for the energy $E$ and the $z$-component of total angular momentum $J_z$ for a pole-dipole-quadrupole body turn out to be
\bea\label{energy_TD}
E_{| \rm TD} &=& \frac{p_t \left(M a S - \mu r^3 \right) + M S p_{\phi}}{\mu\, r^3},\\\label{momentum_TD}
J_{z | \rm TD} &=& \frac{p_t \left(a^2 M - r^3 \right) S + p_\phi \left(MaS + \mu\, r^3 \right)}{\mu\, r^3}.
\eea
The time and azimuthal covariant components of the four-momentum are substituted by their contravariant counterparts, i.e., $p_{\alpha}=g_{\alpha\beta}\,p^\beta$. Substituting Eq.~\eqref{eq:ptW} and $p^{\phi}=W p^t$ into Eqs.~\eqref{energy_TD} and \eqref{momentum_TD}, we find the expressions for energy $E$ and the $z$-component of the total angular momentum $J_z$ in terms of $r, M, a, S$ and $W$, given by
\beq\label{eq:EE-TD}
E_{| \rm TD} = \frac{M S \left(W \left(a^2+r^2\right)-a\right) + \mu  r^2 (r + 2 M (a W-1))}{r^{5/2} \sqrt{r -2 M (a W-1)^2 - r W^2 \left(a^2+r^2\right)}},
\eeq
\bea\non
J_{z | \rm TD} = \frac{\mu  r^2 \left(a^2 W (2 M+r)-2 a M+r^3 W\right)}{r^{5/2} \sqrt{r -2 M (a W-1)^2 - r W^2 \left(a^2+r^2\right)}}\\\label{eq:Jz-TD} +\frac{ \left(a M W \left(a^2+3 r^2\right)-a^2 M-2 M r^2+r^3\right) S}{r^{5/2} \sqrt{r -2 M (a W-1)^2 - r W^2 \left(a^2+r^2\right)}}.
\eea
The equations (\ref{eq:EE-TD}) and (\ref{eq:Jz-TD}) are without any approximation in spin $S$. In order to find the energy $E$ and $J_z$ in terms of orbital parameters ($r, M, a, S$), we need $W$ as a function of orbital parameters. However, we can not find $W$ analytically, but it can be found numerically (without any approximation) using a system of equations (\ref{eq:omega_TD}) and (\ref{eq:W_TD}).
The analytical expression for $W_{\pm}$ in terms of $r, a, S, k$, up to $O(\sigma^6)$ is given in the Appendix. The $+$ sign corresponds to the co-rotation, while the $-$ sign signifies the counter-rotation, with respect to the $z$-component of the total angular momentum $J_z$. We wish to stress at this point that the positive orbital frequencies $\Omega_{+}$ (and $W_{+}$) are associated with $E>0$ and $J_z>0$, while the negative orbital frequencies $\Omega_{-}$ (and $W_{-}$) are associated with $E>0$ and $J_z<0$. The ISCOs can be found by solving $\frac{\d E}{\d r}=0$ or $\frac{\d J_z}{\d r}=0$. Since $E$ and $J_z$ are functions of $W$, which itself is a function of radius $r$, and we do not have $W$ in an analytical form, we cannot directly compute $\frac{\d E}{\d r}$ or $\frac{\d J_z}{\d r}$. Therefore, it is impossible to determine the ISCOs analytically or numerically without making approximations in spin $\sigma$, under TD SSC. However, using the analytical expression of $W$ with a spin approximation up to $O(\sigma^6)$, we can obtain approximate ISCOs under TD SSC. A detailed analysis of ISCOs under TD SSC is presented in the Appendix \ref{sec:ISCOsMP-TD}.

\subsubsection{ISCOs under MP SSC} \label{sec:ISCO_MP} 

To find the ISCOs under MP SSC, we replace the reference four-vector $V^\alpha=u^\alpha$ in Eqs.~\eqref{eq:energy} and \eqref{eq:Jz}, the expressions for the energy $E$ and the $z$-component of total angular momentum $J_z$ for a pole-dipole-quadrupole body turn out to be
\bea
E_{|\rm MP} & = & -p_t + \frac{M  \left(a u_t + u_\phi \right)S}{r^3}, \label{eq:EMP}\\\
J_{z | \rm MP} & = & p_\phi + \frac{\left[u_t (M a^2 - r^3) + M a u_\phi \right] S}{r^3}. \label{eq:JzMP}
\eea
The time and azimuthal covariant components of the four-momentum and four-velocity are substituted by their contravariant counterparts, i.e., $p_{\alpha}=g_{\alpha\beta}\,p^\beta$, and $u_{\alpha}=g_{\alpha\beta}\,u^\beta$, where $p^t$ and $p^\phi$ are found using Eqs.~\eqref{eq:MP_Str} and \eqref{eq:MP_pr}. These constraints, combined with the definition of orbital frequency $\Omega=u^\phi/u^t$, and a relation \eqref{eq:utOmega}, provide the expression for energy $E$ and $z$-component of total angular momentum $J_z$ as a function of $r, a, k$ and $\Omega$. The resulting expressions for energy $E_{|\rm MP}$ and $z$-component of total angular momentum $J_{z |\rm MP}$ under MP SSC, as a function of $r, a, S, k$ and $\Omega$, without truncation in the spin parameter $S$, are presented in Eqs.~\eqref{E_MP} and \eqref{Jz_MP}, respectively (Appendix \ref{sec:LenghtExp}).
Similar to the TD SSC, we can use $\frac{\d E}{\d r}=0$ or $\frac{\d Jz}{\d r}=0$ to find the ISCOs under MP SSC. In contrast to the geodesics or pole-dipole case, the ISCOs calculated using the functions $E$ and $J_z$ in the pole-dipole-(spin-induced)quadrupole are inconsistent. This inconsistency arises because the mass $m$ of the spin-induced quadrupole body is not conserved in the underlying system -- it varies from orbit to orbit. As a result, the ISCOs derived from the $E$ and $J_z$ functions do not agree. We determine the ISCOs using both functions $E$ and $J_z$ and make a comparison. A detailed description is given in the Appendix \ref{sec:ISCOsE-Jz}. In order to make a comparison of ISCOs under MP SSC with those under TD SSC, we also analytically compute the ISCOs under MP SSC using the approximate orbital frequency $\Omega_{\rm MP}$ expanded up to $O(\sigma^6)$. A detailed comparison is given in Appendices~\ref{sec:ISCOsMP-TD}.

\section{Discussion and Conclusions} \label{sec:Conc}

We have analyzed the finite-size effects on the dynamics of extended test bodies within the pole-dipole-(spin-induced)quadrupole MPD framework moving on an equatorial plane of a Kerr BH. We have derived the analytical expressions for the orbital frequency of such an extended test body, under both the TD and the MP SSCs. To examine the discrepancies in the orbital frequency under these SSCs, we applied a power series expansion analysis with respect to the spin magnitude as done in~Ref.~\cite{Tim-etal:2023PhRvD:}. Our findings reveal that, in analogy to the pole-dipole approximation, the orbital frequencies (prograde and retrograde) under the TD and MP SSCs are equivalent up to the quadratic order in spin $\sigma$, i.e., the terms up to $\mathcal{O}(\sigma^2)$ are identical.

Having the orbital frequencies expanded as power series in spin $\sigma$, we shift the centroid of the extended body from the MP SSC frame to the TD SSC frame. The inclusion of centroid position corrections leads to an improvement in convergence, enhancing the agreement by one order in spin. Specifically, when the centroid position corrections are applied, the orbital frequencies under the TD and MP SSCs are equivalent up to the cubic order in spin $\sigma$. This is the same result that was obtained in Ref.~\cite{Tim-etal:2023PhRvD:} for the pole-dipole approximation. However, for $\hat{k}=1$, the convergence increases to the quartic order in spin $\sigma$. It is noticeable that the spin corrections due to the SSC shift start from $\mathcal{O}(\sigma)^4$, i.e. $\tilde{\sigma}=\sigma+\mathcal{O(\sigma)}^4$, but for $\hat{k}=1$, the $\mathcal{O}(\sigma)^4$ vanishes and the correction in spin takes the form $\tilde{\sigma}=\sigma+\mathcal{O}(\sigma)^5$. Hence, for BHs ($\hat{k}=1$), the inclusion of the (spin-induced)quadrupole improved the convergence between the circular equatorial orbital frequencies of the MP frame and the TD frame by one order in spin with respect to the pole-dipole approximation. This improvement came without performing the quadrupole term transformation due to the SSC change. We leave it to a future study to show whether this is a drawback in our approach. We are expected to be able to discern effects up to $\sigma^2$ in the gravitational waves emitted by EMRIs \cite{Piovano:2021PhRvD,Rahman-Bhattacharyya:2023PhRvD:}. Hence, our study indicates that the SSC choice is not important for gravitational wave modeling. 

Our attempt to find the ISCO in the pole-dipole-(spin-induced)quadrupole formalism under TD SSC and MP SSC could not be fully achieved. Even if for the MP SSC in Ref.~\cite{Tim-etal:2023PhRvD:}, the orbital frequency has been found analytically, but for the TD SSC, the frequency, along with the function $W$ could not be found analytically. Apart from this issue, the major problem is that the masses in the pole-dipole-(spin-induced)quadrupole framework are not conserved quantities under the TD and MP SSCs. The fact that they are constant for a circular equatorial orbit allowed us to find the respective frequencies, but the mass changes from one circular equatorial orbit to another. Hence, in our energy and angular momentum formulae, the masses $\mu$ and $m$ are not constants anymore, but functions of $r$. This is a fact that we cannot take into account self-consistently in our calculations.

\section*{Acknowledgments}

M.S. has been supported by GA\v{C}R-25-15272I, G.L.G. has been supported by the fellowship Lumina Quaeruntur No. LQ100032102 of the Czech Academy of Sciences, and M.K. has been supported by the Institute of Physics, Silesian University in Opava. We would like to thank Filipe Costa and Abraham Harte for their constructive criticism.


\def\prc{Phys. Rev. C}
\def\pre{Phys. Rev. E}
\def\prd{Phys. Rev. D}
\def\jcap{Journal of Cosmology and Astroparticle Physics}
\def\apss{Astrophysics and Space Science}
\def\mnras{Monthly Notices of the Royal Astronomical Society}
\def\apj{The Astrophysical Journal}
\def\aap{Astronomy and Astrophysics}
\def\actaa{Acta Astronomica}
\def\pasj{Publications of the Astronomical Society of Japan}
\def\apjl{Astrophysical Journal Letters}
\def\pasa{Publications Astronomical Society of Australia}
\def\nat{Nature}
\def\physrep{Physics Reports}
\def\araa{Annual Review of Astronomy and Astrophysics}
\def\apjs{The Astrophysical Journal Supplement}
\def\aapr{The Astronomy and Astrophysics Review}
\def\procspie{Proceedings of the SPIE}

\bibliographystyle{unsrt}
\bibliography{reference}

\appendix

\begin{figure*}[ht!]
\begin{center}
\includegraphics[width=\hsize]{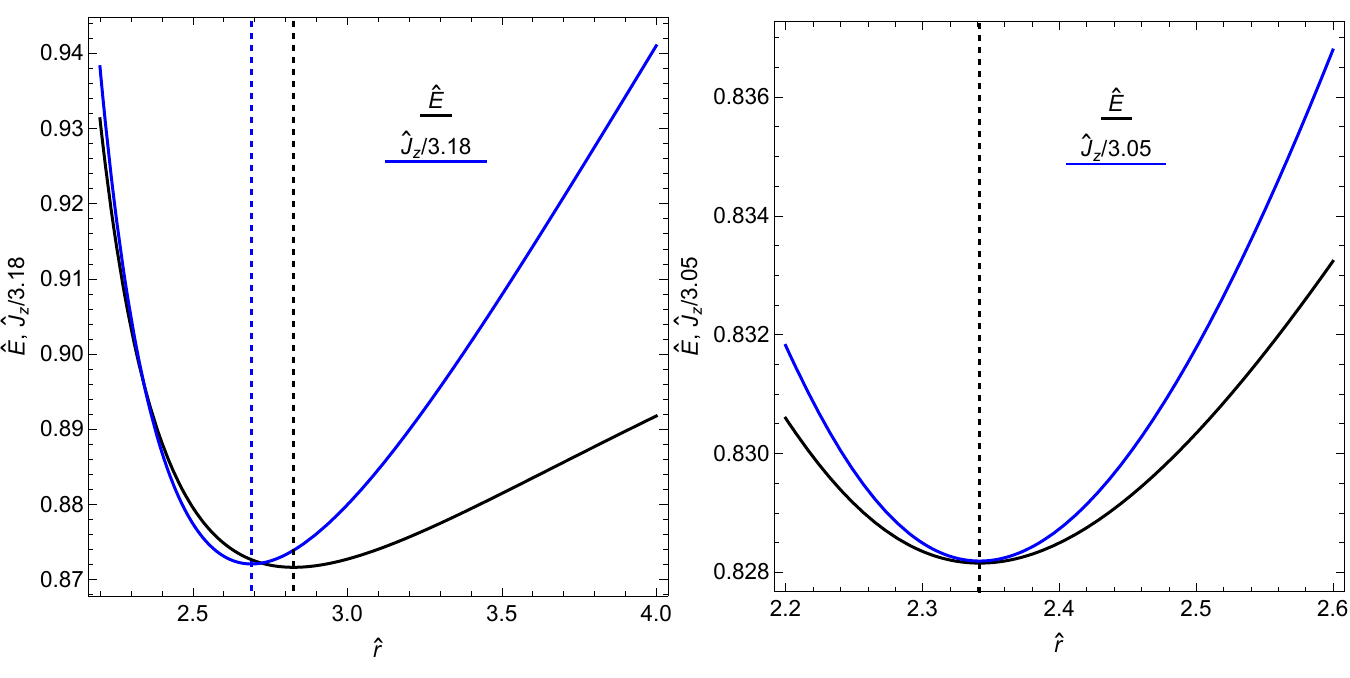}
\end{center}
\caption{Left panel: Energy $\hat{E}$ (black curve) and rescaled $z$-component of the total angular momentum $\hat{J_z}$ (blue curve) of a pole-dipole-(spin-induced)quadrupole body under MP SSC with $\sigma=0.7,~\hat{k}=1$ moving on circular equatorial orbits in a Kerr background with $\hat{a}=0.7$. The vertical black and blue dashed lines indicate the positions of ISCOs, corresponding to the minima of $\hat{E}$ and $\hat{J_z}$, respectively. Right panel: Energy $E$ (black curve) and rescaled $z$-component of the total angular momentum $\hat{J_z}$ (blue curve) of a pole-dipole body under MP SSC with $\sigma=0.7,~\hat{k}=0$ moving on circular equatorial orbits in a Kerr background with $\hat{a}=0.7$. The vertical black, the position of ISCO, corresponding to the identical minima of $\hat{E}$ and  $\hat{J_z}$}
\label{fig:E_Jz}
\end{figure*}

\section{Lengthy expressions}\label{sec:LenghtExp}

This section contains lengthy expressions. The following frequency power series expansion components refer to Tab.~\ref{tab:CEOFrp}
\begin{widetext}
\bea
\hat{\Omega}_{2}(\hat{r},\hat{a},\hat{k}) =\frac{3 \left(\hat{a}\mp \sqrt{\hat{r}}\right) \left( \mp 9 \hat{a}^2 \mp 7 \hat{r}^2 -\hat{a} \sqrt{\hat{r}} \left(3 \hat{r}+1\right)\right)}{8 \hat{r}^{5/2} \left(\hat{a}\pm \hat{r}^{3/2}\right)^3} 
+ \frac{3 \hat{k} \left( \pm 5 \hat{a}^2 - 4 \hat{a} \sqrt{\hat{r}} \pm \left(\hat{r}-2\right) \hat{r}\right)}{4 \hat{r}^{5/2} \left(\hat{a}\pm \hat{r}^{3/2}\right)^2}.
\eea
\bea\non
\hat{\Omega}_{3,\text{TD}}(\hat{r},\hat{a},\hat{k}) &=& \frac{3 \left(\hat{a} \mp \sqrt{\hat{r}}\right) \left( \pm 45 \hat{a}^4  \pm 8 \hat{r}^4 + \hat{a} \left(9 \hat{r}-26\right) \hat{r}^{5/2} + 3 \hat{a}^3 \sqrt{\hat{r}} \left(12 \hat{r}-1\right)\pm \hat{a}^2 \hat{r} \left(9 \hat{r}^2+42 \hat{r}-16\right)\right)}{16 \hat{r}^{9/2} \left(\hat{a}\pm \hat{r}^{3/2}\right)^4}\\ 
&+&
\frac{3 \hat{k} \left( \mp 45 \hat{a}^4 \mp 2 \hat{r}^3 \left(\hat{r}+4\right) - \hat{a}^3 \sqrt{\hat{r}} \left(15 \hat{r}-44\right)\mp \hat{a}^2 \hat{r} \left(19 \hat{r}-26\right) -\hat{a} \left(3 \left(\hat{r}-14\right) \hat{r}+20\right) \hat{r}^{3/2} \right)}{8 \hat{r}^{9/2} \left(\hat{a}\pm \hat{r}^{3/2}\right)^3}.
\eea
\bea\non
\hat{\Omega}_{3,\text{MP}}(\hat{r},\hat{a},\hat{k}) &=& \frac{3 \left(\hat{a} \mp \sqrt{\hat{r}}\right)}{16 \hat{r}^{9/2} \left(\left(\hat{r}-3\right) \sqrt{\hat{r}} \pm 2 \hat{a} \right) \left(\hat{a}\pm \hat{r}^{3/2}\right)^4}
[\hat{a}^3 \hat{r} \left(3 \hat{r} \left(18 \hat{r}+7\right)-23\right)+\hat{a} \hat{r}^3 \left(\hat{r} \left(9 \hat{r}+11\right)-18\right)+90 \hat{a}^5 \\\non & \pm& 
117 \hat{a}^4 \left(\hat{r}-1\right) \sqrt{\hat{r}}\mp 8 \hat{r}^{9/2} \left(9-4 \hat{r}\right)\mp \hat{a}^2 \hat{r}^{5/2} \left(9 \hat{r}^2+57 \hat{r}-170\right)] -\frac{3 \hat{k}}{8 \hat{r}^{9/2} \left(\left(\hat{r}-3\right) \sqrt{\hat{r}} \pm 2 \hat{a} \right) \left(\hat{a}\pm \hat{r}^{3/2}\right)^3}
\\\non &\times& 
\left[\hat{a} \left(\hat{r} \left(\hat{r} \left(3 \hat{r}-35\right)+126\right)-36\right) \hat{r}^2 + \hat{a}^3 \hat{r} \left(3 \hat{r} \left(5 \hat{r}-13\right)+68\right)\mp \hat{a}^2 \hat{r}^{3/2} \left(\hat{r} \left(25 \hat{r}-167\right)+94\right)  \pm 2 \left(\hat{r}-5\right)  \right. \\ &\times& \left. \hat{r}^{9/2} + 90 \hat{a}^5 \mp \hat{a}^4 \sqrt{\hat{r}} \left(75 \hat{r}-211\right) \right].
\eea
\bea\non
\hat{\Omega}_{4,\text{TD}}(\hat{r},\hat{a},\hat{k}) &=& \frac{3 \left(\hat{a}\mp \sqrt{\hat{r}}\right)}{128 \hat{r}^{13/2} \left(\hat{a}\pm \hat{r}^{3/2}\right)^5} 
[3 \hat{a} \left(17 \hat{r}+149\right) \hat{r}^{9/2} -\hat{a}^3 \left(9 \hat{r} \left(\hat{r} \left(15 \hat{r}+67\right)-145\right)+67\right) \hat{r}^{3/2} - 9 \hat{a}^5 \sqrt{\hat{r}} \left(141 \hat{r}-23\right) \\\non &\mp& 
3 \hat{a}^2 \hat{r}^3 \left(45 \left(\hat{r}-1\right) \hat{r}-59\right)\mp 3 \hat{a}^4 \hat{r} \left(3 \hat{r} \left(75 \hat{r}+101\right)-191\right) \mp 945 \hat{a}^6 \mp 13 \hat{r}^6 ] 
\\\non &+& \frac{3 \hat{k}}{32 \hat{r}^{13/2} \left( \pm 2 \hat{a} + \left(\hat{r}-3\right) \sqrt{\hat{r}}\right) \left(\hat{a}\pm \hat{r}^{3/2}\right)^4}
[-2 \hat{a} \left(\hat{r} \left(27 \hat{r}-268\right)+156\right) \hat{r}^4 + 1350 \hat{a}^7 \mp 9 \hat{a}^6 \sqrt{\hat{r}} \left(195 \hat{r}-409\right) \\\non &\mp& \hat{r}^{11/2} \left(\left(\hat{r}-53\right) \hat{r}+174\right) + 2 \hat{a}^5 \hat{r} \left(3 \hat{r} \left(135 \hat{r}-397\right)+793\right)\mp \hat{a}^4 \hat{r}^{3/2} \left(\hat{r} \left(3 \hat{r} \left(45 \hat{r}-62\right)-3088\right)+2597\right) 
\\\non &+& 
\left(2 \hat{a}^3 \hat{r}^2 \left(\hat{r} \left(\hat{r} \left(81 \hat{r}-919\right)+2805\right)-1006\right)\mp \hat{a}^2 \hat{r}^{5/2} \left(\hat{r} \left(\hat{r} \left(\hat{r} \left(27 \hat{r}-296\right)+800\right)-1583\right)+186\right)\right)]
\\\non &-&
\frac{9 \hat{k}^2}{32 \hat{r}^{13/2} \left(\left(\hat{r}-3\right)\sqrt{\hat{r}} \pm 2 \hat{a} \right) \left(\hat{a}\pm \hat{r}^{3/2}\right)^3} [2 \hat{a}^2 \left(\hat{r} \left(\hat{r} \left(5 \hat{r}-66\right)+214\right)-148\right) \hat{r}^2 + 150 \hat{a}^6 \mp 25 \hat{a}^5 \sqrt{\hat{r}} \left(5 \hat{r}-17\right) 
\\\non &+& 
\hat{r}^4 \left(\left(\hat{r}-8\right) \hat{r} \left(\hat{r}+1\right)+52\right)\pm \hat{a} \hat{r}^{5/2} \left(5 \hat{r}-14\right) \left(\hat{r} \left(\hat{r}+13\right)-2\right)+ \hat{a}^4 \hat{r} \left(5 \hat{r} \left(5 \hat{r}-27\right)+168\right) \pm 2 \hat{a}^3 \hat{r}^{3/2} 
\\ &\times& \left(\hat{r} \left(25 \hat{r}-139\right)+190\right)].
\eea
\bea\non
\hat{\Omega}_{4,\text{MP}}(\hat{r},\hat{a},\hat{k}) &=& \frac{3}{128 \hat{r}^{13/2} \left( \pm 2 \hat{a} + \left(\hat{r}-3\right) \sqrt{\hat{r}}\right)^2 \left(\hat{a} \pm \hat{r}^{3/2}\right)^5}[\hat{a}^4 \left(\hat{r} \left(\hat{r} \left(9 \hat{r} \left(3 \hat{r} \left(22-5 \hat{r}\right)+239\right)-7540\right)+7582\right)+1500\right) \hat{r}^{5/2} 
\\\non &+&
\hat{a}^2 \left(\hat{r} \left(\hat{r} \left(90 \hat{r}^2+92 \hat{r}-4991\right)+12978\right)-1593\right) \hat{r}^{9/2} + \hat{a}^6 \hat{r}^{3/2} \left(9 \hat{r} \left(\hat{r} \left(1330-501 \hat{r}\right)+155\right)+2840\right) \\\non &+&
36 \hat{a}^8 \left(379-246 \hat{r}\right) \sqrt{\hat{r}} \mp \hat{a} \hat{r}^6 \left(\hat{r} \left(\hat{r} \left(640 \hat{r}-3541\right)+4098\right)+567\right)\mp \hat{a}^5 \hat{r}^2 \left(\hat{r} \left(27 \hat{r} \left(3 \hat{r} \left(46-15 \hat{r}\right)+377\right) 
\right. \right. \\\non &-& \left. \left. 25684\right)+970\right)\mp \hat{a}^3 \hat{r}^3 \left(\hat{r} \left(\hat{r} \left(-342 \hat{r}^2+5300 \hat{r} - 18299\right)+6834\right)-549\right) + \hat{r}^{15/2} \left(5 \hat{r} \left(137 \hat{r}-534\right)+2421\right) 
\\\non &\mp&
9 \hat{a}^7 \hat{r} \left(\hat{r} \left(969 \hat{r}-2258\right)+1625\right)+\hat{a}^9 \mp 3780] + 
\frac{3 \hat{k}}{32 \hat{r}^{13/2} \left(\pm 2 \hat{a} +\left(\hat{r}-3\right) \sqrt{\hat{r}}\right)^2 \left(\hat{a}\pm \hat{r}^{3/2}\right)^4}
[\pm 2700 \hat{a}^8
\\\non & + & 
4 \hat{a} \hat{r}^{9/2} \left(\hat{r} \left(\left(597-62 \hat{r}\right) \hat{r}-1383\right)+522\right) + 36 \hat{a}^7 \sqrt{\hat{r}} \left(135 \hat{r}-281\right)\mp \hat{a}^6 \hat{r} \left(15 \hat{r} \left(225 \hat{r}-722\right)+10519\right)
\\\non &+& 
4 \hat{a}^5 \hat{r}^{3/2} \left(3 \hat{r} \left(6 \hat{r} \left(15 \hat{r}-44\right)-287\right)+277\right)\mp \hat{r}^6 \left(\hat{r} \left(\hat{r} \left(23 \hat{r}-136\right)-69\right)+666\right) + 4 \hat{a}^3 \hat{r}^{5/2} \left(\hat{r} \left(\hat{r} \left(54 \hat{r}^2 
\right. \right. \right. \\\non &-& \left. \left. \left.
417 \hat{r}+1369\right)-1974\right)+258\right) \mp \hat{a}^4 \hat{r}^2 \left(\hat{r} \left(\hat{r} \left(15 \hat{r} \left(9 \hat{r}+11\right)-7622\right)+19865\right)-5743\right) \mp \hat{a}^2 \hat{r}^3 \left(\hat{r} \left(\hat{r} \left(\hat{r} \right. \right. \right. \\\non &\times& \left. \left. \left.
\left(\hat{r} \left(27 \hat{r}-17\right)-1996\right)+6905\right)-4593\right)+450\right)]
- \frac{9 \hat{k}^2 \left(5 \hat{a}^2 \mp 4 \hat{a} \sqrt{\hat{r}} + \left(\hat{r}-2\right) \hat{r} \right) }{32 \hat{r}^{13/2} \left(\left(\hat{r}-3\right)\sqrt{\hat{r}} \pm 2 \hat{a} \right) \left(\hat{a}\pm \hat{r}^{3/2}\right)^3} 
\left[30 \hat{a}^4 \mp 5 \hat{a}^3 \right. \\ &\times& \left. 
\sqrt{\hat{r}} \left(5 \hat{r}-9\right) + \hat{a}^2 \hat{r} \left(\hat{r} \left(5 \hat{r}+3\right)-16\right)\mp \hat{a} \hat{r}^{3/2} \left(\hat{r} \left(9 \hat{r}-47\right)+26\right)+\left(\left(\hat{r}-5\right) \hat{r}+14\right) \hat{r}^3
\right].
\eea
The expressions for $\hat{\Omega}_{4,\text{TD}}^{'}(\hat{r},\hat{a},\hat{k})$ and $\hat{\Omega}_{4,\text{TD}}^{''}(\hat{r},\hat{a},\hat{k})$ come after shifting the centroid from MP SSC to TD SSC. Taking into account only a radial shift at zero-order estimate, we have
\bea\non
\hat{\Omega}_{4,\text{TD}}^{'}(\hat{r},\hat{a},\hat{k}) &=& \frac{3}{128 \hat{r}^{13/2} \left( \pm 2 \hat{a} + \left(\hat{r}-3\right) \sqrt{\hat{r}}\right)^2 \left(\hat{a}\pm \hat{r}^{3/2}\right)^5} [3780 \hat{a}^9 \mp 12 \hat{a}^8 \sqrt{\hat{r}} \left(738 \hat{r}-1297\right) \mp \hat{r}^{15/2} \left(\left(4782-1069 \hat{r}\right) \hat{r} \right. \\\non &-& \left. 
5301\right) +3 \hat{a}^7 \hat{r} \left(\hat{r} \left(2907 \hat{r}-9014\right)+6859\right)\mp \hat{a}^6 \hat{r}^{3/2} \left(9 \hat{r} \left(\hat{r} \left(501 \hat{r}-2290\right)+1829\right)-5144\right) + \hat{a}^3 \hat{r}^3 (-2331 \\\non &+& \hat{r} \left(\hat{r} \left(2 \left(6566-981 \hat{r}\right) \hat{r}-29317\right)+25230\right) ) +\hat{a} \hat{r}^6 \left(\hat{r} \left(\hat{r} \left(1792 \hat{r} 
- 11605\right)+20418\right)-8073\right) \mp \hat{a}^2 \hat{r}^{9/2} (\hat{r} \left(\hat{r} \right. \\\non &\times& \left. \left(-858 \hat{r}^2 + 9124 \hat{r}-25729\right)+23310\right)-7047) + \hat{a}^5 \hat{r}^2 \left(\hat{r} \left(3 \hat{r} \left(\hat{r} \left(405 \hat{r}-2842\right)+2303\right)+24532\right)-9802\right) 
 \\\non &\mp& \hat{a}^4 \hat{r}^{5/2} (\hat{r} \left(\hat{r} \left(3 \hat{r} \left(\hat{r} \left(45 \hat{r}-518\right)+499\right)+19060\right)-40990\right)+8484)] 
 + 4 \hat{k} [- 2700 \hat{a}^9 \mp 12 \hat{a}^8 \left(923-630 \hat{r}\right) \sqrt{\hat{r}} 
 \\\non &+&  \hat{a}^7 \hat{r} \left(3 \left(8102-2745 \hat{r}\right) \hat{r}-13495\right) \hat{r}^{3/2} \mp \hat{a}^6 (\hat{r} \left(3 \left(6106-1485 \hat{r}\right) \hat{r}-16003\right)+44) + \hat{a} \hat{r}^6 \left(\hat{r} \left(\left(1780-351 \hat{r}\right) 
\right. \right. \\\non &\times& \left. \left.
 \hat{r}-2559\right)+1566\right)\mp \hat{r}^{15/2} \left(\hat{r} \left(\left(328-23 \hat{r}\right) \hat{r}-987\right)+774\right) 
 + \hat{a}^3 \hat{r}^3 \left(\hat{r} \left(\hat{r} \left(\hat{r} \left(\left(2837-243 \hat{r}\right) \hat{r}-12696\right) \right. \right. \right. \\\non &+& \left. \left. \left. 24799\right)-12471\right)+990\right)  +\hat{a}^5 \left(\hat{r} \left(\hat{r} \left(3 \hat{r} \left(1801-405 \hat{r}\right)+2522\right)-20397\right)+10159\right) \hat{r}^2  \mp \hat{a}^2 \hat{r}^{9/2} (\hat{r} \left(\hat{r} \left(\hat{r} 
 \right. \right. \\\non &\times& \left. \left.
 \left(\left(401-27 \hat{r}\right) \hat{r}-2364\right)+6067\right)-8019 \right)+1782)  \mp \hat{a}^4 \hat{r}^{5/2} (\hat{r} \left(\hat{r} \left(\hat{r} \left(9 \hat{r} \left(11-15 \hat{r}\right)+7466\right)-31101\right) \right. \\\non &+& \left. 30343\right) - 6024)]
 + 12 \hat{k}^2 \left(\hat{a}\pm \hat{r}^{3/2}\right)^2 [300 \hat{a}^7 \mp 100 \hat{a}^6 \sqrt{\hat{r}} \left(4 \hat{r}-13\right) \pm \hat{r}^{9/2} \left(\hat{r}-3\right) \left(\left(\hat{r}-8\right)  \left(\hat{r}+1\right) \hat{r} + 52\right) \\\non
 &+& \hat{a}^5 \hat{r} \left(5 \hat{r} \left(35 \hat{r}-214\right)+1611\right)\pm \hat{a}^4 \hat{r}^{3/2} \left(\hat{r} \left(5 \hat{r} \left(5 \hat{r}-22\right)+17\right)+256\right) + 2 \hat{a}^3 \hat{r}^2 \left(\hat{r} \left(\hat{r} \left(35 \hat{r}-346\right)+1035\right) \right. \\\non &-& \left. 
 866\right) + 2 \hat{a}^3 \hat{r}^2 \left(\hat{r} \left(\hat{r} \left(35 \hat{r}-346\right)+1035\right) - 866\right) +2 \hat{a}^3 \hat{r}^2 \left(\hat{r} \left(\hat{r} \left(35 \hat{r}-346\right) + 1035\right)-866\right) \pm 2  \hat{a}^2 \hat{r}^{5/2} (\hat{r} \left(\hat{r} \right. \\ &\times&
\left. \left(\hat{r} \left(5 \hat{r}-76\right)+463\right)-982\right)+472) +\hat{a} \hat{r}^3 \left(\hat{r} \left(\hat{r} \left(\hat{r} \left(7 \hat{r}+22\right)-361\right)+708\right)-84\right)].
\eea
The expressions for both $\hat{\Omega}_{4,\text{TD}}^{'}(\hat{r},\hat{a},\hat{k})$, where we apply the radial first-order shift, and the $\hat{\Omega}_{4,\text{TD}}^{''}(\hat{r},\hat{a},\hat{k})$, where the shift in spin is incorporated as well, can be found in the supplementary material~\cite{SupMat} along with all the $\hat{\Omega}_n(\hat{r},\hat{a},\hat{k})$ including $\hat{\Omega}_5$, $\hat{\Omega}_5^{'}$,  $\hat{\Omega}_5^{''}$, $\hat{\Omega}_6$, $\hat{\Omega}_6^{'}$. The expressions for $\hat{W}_n$, where $n=0,1,2,3,4$ are presented below, but also in the supplementary material~\cite{SupMat}, where one can find the expressions for $n=5,6$ as well
\beq
\hat{W}_0 (\hat{r}, \hat{a}) = \frac{1}{\hat{a} \pm \sqrt{\hat{r}^{3}}},
\eeq
\beq
\hat{W}_1 (\hat{r}, \hat{a}) = \frac{3 \left(\pm \hat{a}-\sqrt{\hat{r}}\right)}{2 \sqrt{\hat{r}} \left(\hat{a} \pm \hat{r}^{3/2}\right)^2},
\eeq
\beq
\hat{W}_2 (\hat{r}, \hat{a},\hat{k}) = \frac{3 \left(\hat{a}\mp \sqrt{\hat{r}}\right) \left(\mp 9 \hat{a}^2 \pm \hat{r}^2-\hat{a} \sqrt{\hat{r}} \left(3 \hat{r}-7\right)\right)}{8 \hat{r}^{5/2} \left(\hat{a}\pm \hat{r}^{3/2}\right)^3} 
+ \frac{3 \hat{k} \left(\pm 5 \hat{a}^2  - 8 \hat{a} \sqrt{\hat{r}}\pm \hat{r} \left(\hat{r}+2\right)\right)}{4 \hat{r}^{5/2} \left(\hat{a}\pm \hat{r}^{3/2}\right)^2}.
\eeq
\bea\non
\hat{W}_3 (\hat{r}, \hat{a},\hat{k}) &=& \frac{3 \left(\hat{a}\mp \sqrt{\hat{r}}\right) \left( \pm 45 \hat{a}^4  \mp 16 \hat{r}^4 + \hat{a} \left(9 \hat{r}-26\right) \hat{r}^{5/2} + 3 \hat{a}^3 \sqrt{\hat{r}} \left(12 \hat{r}-17\right)\pm \hat{a}^2 \hat{r} \left(9 \hat{r}^2-6 \hat{r}+8\right)\right)}{16 \hat{r}^{9/2} \left(\hat{a}\pm \hat{r}^{3/2}\right)^4} \\ 
&+&
\frac{3 \hat{k} \left( \mp 45 \hat{a}^4 \pm \hat{a}^2 \left(\hat{r}-34\right) \hat{r}  \pm 2 \left(\hat{r}-10\right) \hat{r}^3 -\hat{a} \left(3 \hat{r}^2-34 \hat{r}+8\right) \hat{r}^{3/2}-\hat{a}^3 \sqrt{\hat{r}} \left(15 \hat{r}-88\right)  \right)}{8 \hat{r}^{9/2} \left(\hat{a}\pm \hat{r}^{3/2}\right)^3}. 
\eea
\bea\non
\hat{W}_4 (\hat{r}, \hat{a},\hat{k}) &=& \frac{3 \left(\hat{a} \mp \sqrt{\hat{r}}\right)}{128 \hat{r}^{13/2} \left(\hat{a}\pm \hat{r}^{3/2}\right)^5} \left[\hat{a}^3 \left(13-9 \hat{r} \left(5 \hat{r} \left(3 \hat{r}+7\right)-129\right)\right) \hat{r}^{3/2} \mp 3 \hat{a}^2 \hat{r}^3 \left(45 \hat{r}^2-381 \hat{r}+149\right) \right. \\\non &-& \left.  9 \hat{a}^5 \sqrt{\hat{r}} \left(141 \hat{r}-151\right) \mp 3 \hat{a}^4 \hat{r} \left(3 \hat{r} \left(75 \hat{r}-59\right)+145\right) - 3 \hat{a} \left(59-65 \hat{r}\right) \hat{r}^{9/2}\pm 67 \hat{r}^6  \mp 945 \hat{a}^6 \right] 
\\\non &+& \frac{3 \hat{k}}{32 \hat{r}^{13/2} \left( \pm 2 \hat{a} +\left(\hat{r}-3\right) \sqrt{\hat{r}}\right) \left(\hat{a}\pm \hat{r}^{3/2}\right)^4} [2 \hat{a}^5 \hat{r} \left(3 \hat{r} \left(135 \hat{r}-709\right)+3325\right)  \mp 3 \hat{a}^6 \sqrt{\hat{r}} \left(1739-585 \hat{r}\right) 
\\\non &+&
1350 \hat{a}^7 \mp \hat{r}^{11/2} \left(25 \hat{r}^2-\hat{r}-102\right) + 2 \hat{a} \hat{r}^4 \left(\hat{r} \left(197 \hat{r} - 688\right)+444\right)\mp \hat{a}^4 \hat{r}^{3/2} \left(\hat{r} \left(9 \hat{r} \left(15 \hat{r}-82\right)+668\right) 
\right. \\\non &-& \left. 
2519\right) 
+ 2 \hat{a}^3 \hat{r}^2 \left(\hat{r} \left(\hat{r} \left(81 \hat{r}-1123\right)+2629\right)-362\right)\mp \hat{a}^2 \hat{r}^{5/2} \left(\hat{r} \left(\hat{r} \left(\hat{r} \left(27 \hat{r}-692\right)+3444\right)-4315\right)+462\right)]
\\\non &+& 
\frac{9 \hat{k}^2 \left(5 \hat{a}^2 \mp 8 \hat{a} \sqrt{\hat{r}} + \hat{r} \left(\hat{r}+2\right) \right)}{32 \hat{r}^{13/2} \left( \pm 2 \hat{a} +\left(\hat{r}-3\right) \sqrt{\hat{r}}\right) \left(\hat{a}\pm \hat{r}^{3/2}\right)^3} 
\left[-\hat{a}^2 \hat{r} \left(\hat{r} \left(5 \hat{r}-9\right)+20\right)\mp \hat{a} \hat{r}^{3/2} \left(\hat{r} \left(13 \hat{r}-55\right)+22\right)-30 \hat{a}^4 \right. \\ &\pm& \left. \hat{a}^3 \sqrt{\hat{r}} \left(25 \hat{r}-69\right)-\left(\left(\left(\hat{r}-17\right) \hat{r}+34\right) \hat{r}^3\right) \right].
\eea

The lengthy expressions for energy~\eqref{eq:EMP}  and the angular momentum~\eqref{eq:JzMP} under MP SSC are given by
\bea\non
E_{| \rm MP} &=& \frac{-1}{\chi} [m r^3 (2 M (a \Omega -1)+r) \left(r \left(\Omega ^2 \left(a^2+r^2\right)-1\right)+2 M (a \Omega -1)^2\right) 
+ S  \left(-3 M \Omega ^2 \left(k S \left(2 a^2+r^2\right) \right. \right. \\\non &\times & \left. \left. \left(a^2+r (r-2 M)\right) + a r \left(a^2 (2 M+r)+r^3\right)\right) + 3 a M \Omega  \left(k S \left(a^2+r (r-2 M)\right)+2 a M r\right) + \Omega ^3 \left(2 a^4 M r \right. \right. \\\label{E_MP} &\times& \left. \left.  (M+r)+3 a k M S \left(a^2+r^2\right) \left(a^2+r (r-2 M)\right)+a^2 r^4 (3 M-r)+r^6 (3 M-r)\right)+a M r (r-2 M)\right)],
\eea
\bea\non
J_{z| \rm MP} &=& \frac{1}{\chi} [-3 a^3 k M S^2+\Omega  \left(3 k M S^2 \left(2 a^2+r^2\right) \left(a^2+r (r-2 M)\right)+m r^3 \left(a^2 \left(r^2-12 M^2\right)+r^3 (r-2 M)\right) \right. \\\non  &+& \left. 3 a M r S \left(a^2+3 r^2\right) (r-2 M)\right)-3 a M \Omega ^2 \left(k S^2 \left(a^2+r^2\right) \left(a^2+r (r-2 M)\right)-2 m r^3 \left(a^2 (2 M+r)+r^3\right) \right. \\\non &-& \left. 2 a M r S \left(a^2+3 r^2\right)\right)-r \Omega ^3 \left(a^2 (2 M+r)+r^3\right) \left(m r^2 \left(a^2 (2 M+r)+r^3\right) + a M S \left(a^2+3 r^2\right)\right) 2 a^2 M^2 r S \\ &-& 2 a^2 M r^2 S+6 a k M^2 r S^2-3 a k M r^2 S^2+4 a m M^2 r^3-2 a m M r^4+6 M^2 r^3 S-5 M r^4 S+r^5 S]\label{Jz_MP},
\eea
where $\chi$ is given by
\beq
\chi = r^{7/2} \left(-r \Omega ^2 \left(a^2+r^2\right)-2 M (a \Omega -1)^2+r\right)^{3/2} .
\eeq
\end{widetext}

\section{Comparison of ISCO determination using energy and angular momentum}\label{sec:ISCOsE-Jz}

In this section, we discuss the ISCOs calculated using the energy function $E$ and the $z$-component of the total angular momentum $\hat{J_z}$, under the MP SSC, without truncating $\Omega$, $E$, and $J_z$ in spin $\sigma$. The $\Omega$ corresponds to the non-helical solution of the polynomial equation~\eqref{eq:MP_omega} for the MP SSC, and we refer to this approach as ``analytical''. To compare the minima of $E$ and $\hat{J_z}$ functions, presented in Fig.~\ref{fig:E_Jz}, we rescale $\hat{J_z}$ appropriately. In the left panel of Fig.~\ref{fig:E_Jz}, we show the case $\hat{k}=1$, where the different minima of $E$ and $\hat{J_z}$ lead to non-agreeing $r_\textrm{ISCO}$ calculations. The right panel of Fig.~\ref{fig:E_Jz} shows the pole-dipole case $(\hat{k}=0)$, where the identical minima of $E$ and $\hat{J_z}$ lead to an identical $r_\textrm{ISCO}$ calculation. Hence, in contrast to the pole-dipole case, the minima of the $E$ and $\hat{J_z}$ functions yield inconsistent ISCOs. In the spin-induced quadrupole case, the minima of the $\hat{J_z}$ function lie below that of the $E$ function, when the mass of the body is fixed. This discrepancy arises due to the non-conservation of the mass $m$ of the spinning body, which becomes a function of the radial coordinate $r$ and varies from orbit to orbit. In Ref.~\cite{Steinhoff-Dirk:2012:PhRvD:}, the issue of the non-conservation of the mass led to a definition of another mass quantity, but even then, the authors were pushed to adopt an approximative up to $\sigma^3$ scheme of an effective potential. 

\begin{figure*}[t]
\begin{center}
\includegraphics[width=\hsize]{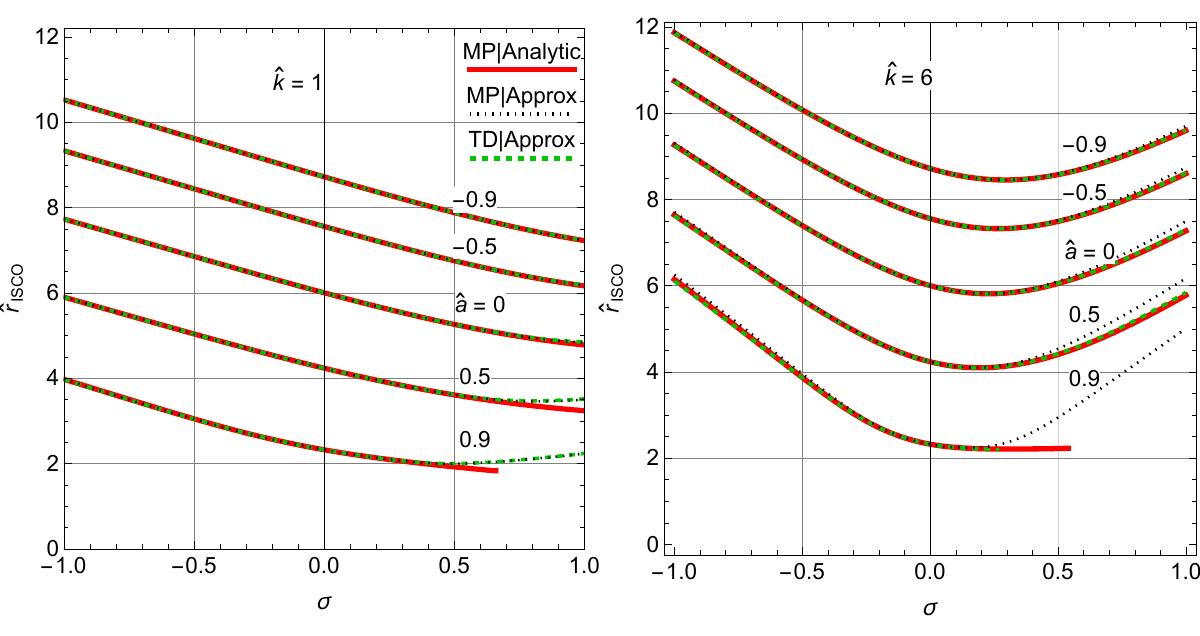}
\end{center}
\caption{ISCOs for three cases (A-C) as a function of spin $\sigma$ for various values of BH spin $\hat{a}=\pm 0.9, \pm0.5, 0$. The first column is plotted for $\hat{k}=1$, while the second column is for $\hat{k}=6$.
The solid red curves correspond to analytical ISCOs under MP SSC without truncation in spin ($r^{\rm Analytic}_{\rm ISCO|MP}$), dashed green curves correspond to approximative MP SSC ($r^{\rm Approx}_{\rm ISCO|MP}$) with spin up to $O(\sigma^6)$, and the dotted black curves correspond to approximative ISCOs under TD SSC $\big(r^{\rm Analytic}_{\rm ISCO|TD}\big)$ with spin up to $O(\sigma^6)$.
}
\label{fig:ISCO}
\end{figure*}

\section{Comparative analysis of ISCOs under TD and MP SSC}\label{sec:ISCOsMP-TD}

\begin{figure*}
\begin{center}
\includegraphics[width=\hsize]{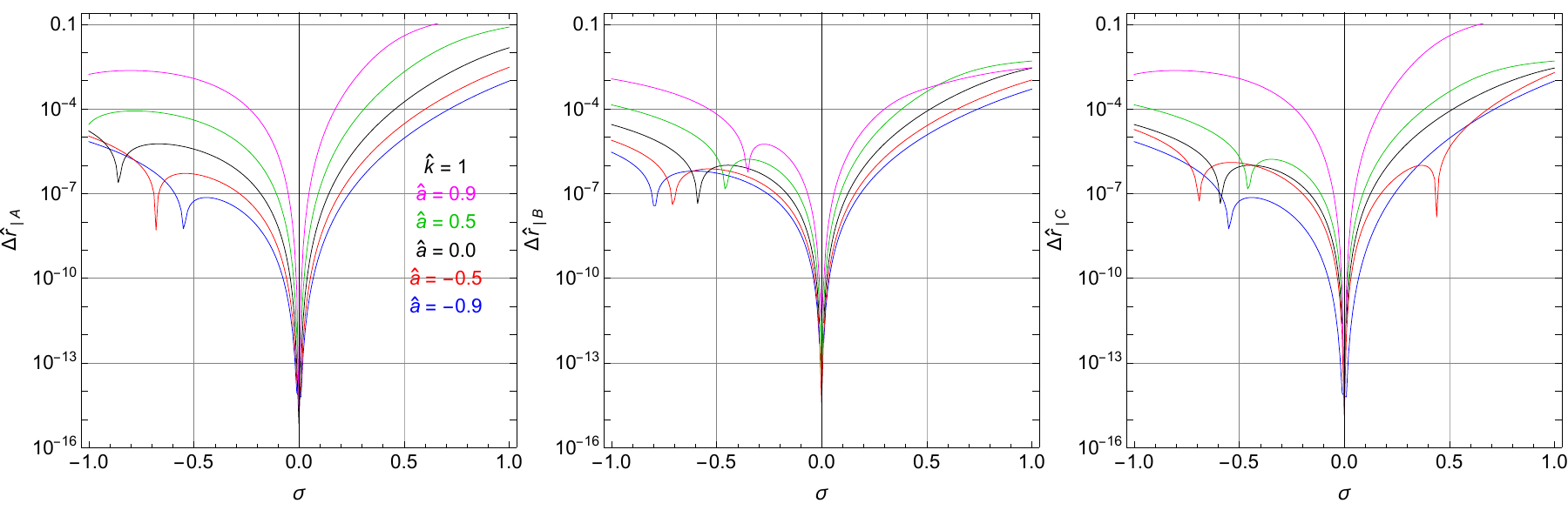}
\includegraphics[width=\hsize]{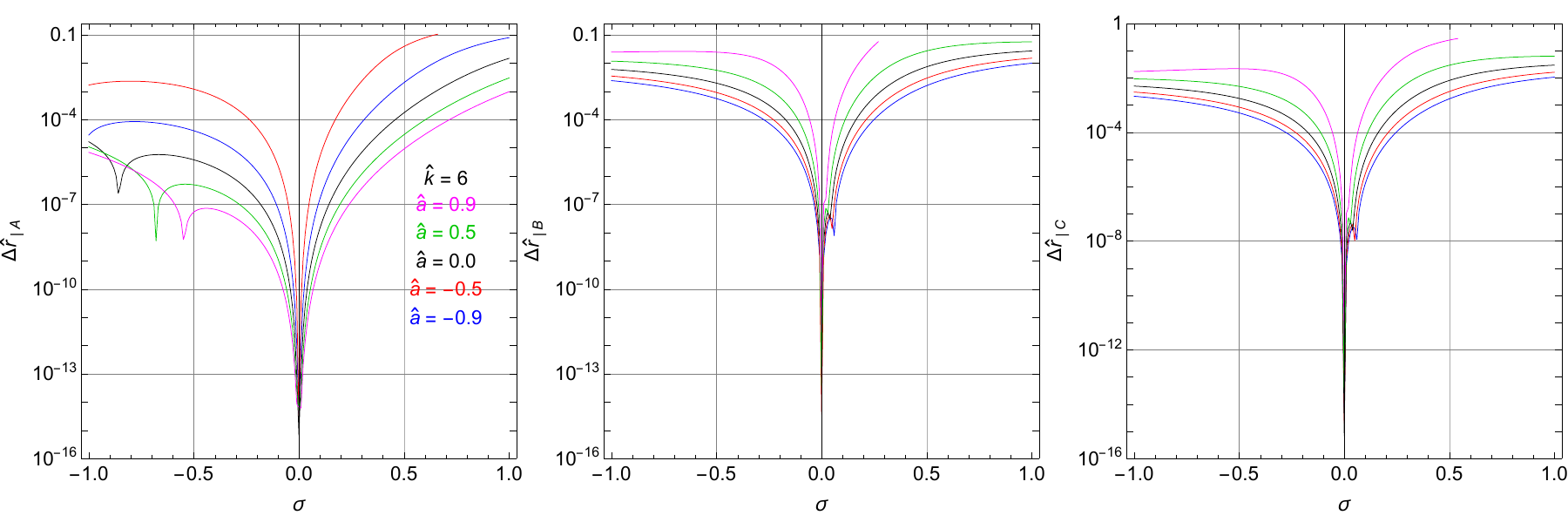}
\end{center}
\caption{Relative difference between ISCO radii for spin-induced quadrupole body orbiting a Kerr BH as a function of spin $\sigma$, for various values of BH spin ($\hat{a}=\pm 0.9, \pm 0.5, 0$) with $\hat{k}=1$ top panels and $\hat{k}=6$ bottom panels, under MP and TD SSC. The first column depicts the relative difference between the analytic and approximate ISCO radii under the MP SSC, i.e., $r^{\rm Analytic}_{\rm ISCO | MP}$ and $r^{\rm Approx}_{\rm ISCO | MP}$. The second column shows the relative difference between the approximate ISCO radii computed using the MP and TD SSCs, i.e., $r^{\rm Approx}_{\rm ISCO|MP}$ and $r^{\rm Approx}_{\rm ISCO|TD}$. The third column depicts relative difference between $r^{\rm Approx}_{\rm ISCO|TD}$ and $r^{\rm Analytic}_{\rm ISCO|MP}$.
}
\label{fig:relatveDiffk1}
\end{figure*}

In this section, we investigate the ISCOs determined by the condition $\frac{\d J_z}{\d r}=0$, under both the TD and the MP SSC. We examine how the ISCOs are affected by the truncation of the orbital frequency in spin $\sigma$, which we call ``approximative'', and the choice of SSC. Specifically, we consider the following three cases
\begin{itemize}
   \item (A) $r^{\rm Analytic}_{\rm ISCO | MP}$: ISCOs under the MP SSC without any truncation of $\hat{J_z}$ in spin $\sigma$.
    \item (B) $r^{\rm Approx}_{\rm ISCO | MP}$: ISCOs under the MP SSC with of $\hat{J_z}$ truncated up to $O(\sigma^6)$.
    \item (C) $r^{\rm Approx}_{\rm ISCO | TD}$: ISCOs under the TD SSC with of $\hat{J_z}$ truncated up to $O(\sigma^6)$.
\end{itemize}
To make a qualitative comparison among these three cases, we compute the relative errors, defined as
\bea\non
\Delta \hat{r}_{|\rm A} = \left|1- \frac{r^{\rm Approx}_{\rm ISCO | MP}}{r^{\rm Analytic}_{\rm ISCO|MP}} \right|,\\\non
\Delta \hat{r}_{|\rm B} = \left|1- \frac{r^{\rm Approx}_{\rm ISCO|MP}}{r^{\rm Approx}_{\rm ISCO | TD}} \right|,\\\label{eq:rel_errors}
\Delta \hat{r}_{|\rm C} = \left|1- \frac{r^{\rm Approx}_{\rm ISCO | TD}}{r^{\rm Analytic}_{\rm ISCO | MP}} \right|.
\eea
Figure~\ref{fig:ISCO} shows the numerical behaviour of the ISCO radii for a pole-dipole-(spin-induced)quadrupole body for the aforementioned three cases, i.e., $r^{\rm Analytic}_{\rm ISCO|MP}$, $r^{\rm Approx}_{\rm ISCO|MP}$, and $r^{\rm Approx}_{\rm ISCO|TD}$. The corresponding relative errors are shown in Fig.~\ref{fig:relatveDiffk1}, where the top panel shows the $\hat{k}=1$ case and the bottom the $\hat{k}=6$ case. We also show the numerical values of ISCO radii for all three considered cases (A-C) for various BH spin ($\hat{a}=-0.9,0,0.9$) and secondary spin values ($\sigma=\pm 0.9, \pm 0.5, \pm 0.1, 0$) in Tab.~\ref{tab:ISCO_combined} for $\hat{k}=1$ and in Tab.~\ref{tab:ISCO_combined_k6} for $k=6$. We can find the approximation ISCO radii for all possible values of BH spin and particle spin ($-1<\hat{a}<1$, $-1<\sigma<1$), under both TD and MP SSC. However, the ISCOs under MP without any truncation of $J_z$ in $\sigma$ could only be found under $\sigma\leq0.653$. We cannot infer wether this has a physical meaning or it is just the limit of our minima finding method. In Fig.~\ref{fig:ISCO}, the green dashed curves coincide visually with the black dotted curves for the case of BHs ($\hat{k}=1$). However, for the case of neutron stars ($\hat{k}=6$), the divergence between ISCO radii increases when the body spins with large $\sigma>0$. Furthermore, for the case of neutron stars, both analytic and approximative ISCOs cannot be found for $\sigma \gtrapprox 0.6$ under MP SSC. To see these discrepancies, we have pushed our models to their limits, for realistic values of the secondary spin $\abs{\sigma}< 0.5$.  Tables~\ref{tab:ISCO_combined} and \ref{tab:ISCO_combined_k6} indicate that these discrepancies are negligible, in particular for BHs.

\begin{table*}[ht]
\centering
\renewcommand{\arraystretch}{1.4}
\resizebox{\textwidth}{!}{%
\begin{tabular}{|c||ccc|ccc|ccc|}
\hline
\multirow{2}{*}{$\sigma$}
& \multicolumn{3}{c|}{$\hat{a} = -0.9$}
& \multicolumn{3}{c|}{$\hat{a} = 0$}
& \multicolumn{3}{c|}{$\hat{a} = 0.9$}
\\ \cline{2-10}
& $r_{\rm ISCO|TD}^{\rm Approx}$ & $r_{\rm ISCO|MP}^{\rm Approx}$ & $r_{\rm ISCO|MP}^{\rm Analytic}$
& $r_{\rm ISCO|TD}^{\rm Approx}$ & $r_{\rm ISCO|MP}^{\rm Approx}$ & $r_{\rm ISCO|MP}^{\rm Analytic}$
& $r_{\rm ISCO|TD}^{\rm Approx}$ & $r_{\rm ISCO|MP}^{\rm Approx}$ & $r_{\rm ISCO|MP}^{\rm Analytic}$
\\ \hline
$-0.9$ & 10.34251 & 10.34250 & 10.34246 & 7.55285 & 7.55272 & 7.55269 & 3.78155 & 3.77839 & 3.78641 \\ \hline
$-0.5$ & 9.61756  & 9.61756  & 9.61757  & 6.85163 & 6.85164 & 6.85166 & 3.04322 & 3.04302 & 3.04662 \\ \hline
$-0.1$ & 8.89511  & 8.89511  & 8.89511  & 6.16523 & 6.16523 & 6.16523 & 2.43595 & 2.43595 & 2.43596 \\ \hline
$0$    & 8.71735  & 8.71735  & 8.71735  & 6.00000 & 6.00000 & 6.00000 & 2.32088 & 2.32088 & 2.32088 \\ \hline
$0.1$  & 8.54178  & 8.54178  & 8.54178  & 5.83895 & 5.83895 & 5.83895 & 2.22045 & 2.22045 & 2.22044 \\ \hline
$0.5$  & 7.87741  & 7.87750  & 7.87743  & 5.25955 & 5.25998 & 5.25907 & 1.99526 & 1.99633 & 1.91959 \\ \hline
$0.9$  & 7.33074  & 7.33279  & 7.32935  & 4.87528 & 4.88406 & 4.84480 & 2.16427 & 2.16939 & -- \\ \hline
\end{tabular}
} 
\caption{Comparison of ISCO radii under TD and MP SSC, for different spin parameters $\hat{a}$ and particle spin values $\sigma$, and $\hat{k}=1$. For \( \hat{a}=0.9 \), the analytic ISCO only exists for \( \sigma \leq 0.653 \); hence it is unavailable for \( \sigma = 0.9 \).}
\label{tab:ISCO_combined}
\end{table*}


\begin{table*}[ht]
\centering
\renewcommand{\arraystretch}{1.4}
\resizebox{\textwidth}{!}{%
\begin{tabular}{|c||ccc|ccc|ccc|}
\hline
\multirow{2}{*}{$\sigma$}
& \multicolumn{3}{c|}{$\hat{a} = -0.9$}
& \multicolumn{3}{c|}{$\hat{a} = 0$}
& \multicolumn{3}{c|}{$\hat{a} = 0.9$}
\\ \cline{2-10}
& $r_{\rm ISCO|TD}^{\rm Approx}$ & $r_{\rm ISCO|MP}^{\rm Approx}$ & $r_{\rm ISCO|MP}^{\rm Analytic}$
& $r_{\rm ISCO|TD}^{\rm Approx}$ & $r_{\rm ISCO|MP}^{\rm Approx}$ & $r_{\rm ISCO|MP}^{\rm Analytic}$
& $r_{\rm ISCO|TD}^{\rm Approx}$ & $r_{\rm ISCO|MP}^{\rm Approx}$ & $r_{\rm ISCO|MP}^{\rm Analytic}$
\\ \hline
$-0.9$ & 11.51749 & 11.49469 & 11.49697 & 8.91591 & 8.86948 & 8.87604 & 5.80209 & 5.66154 & 5.69700 \\ \hline
$-0.5$ & 10.08272 & 10.07718 & 10.07752 & 7.40700 & 7.39269 & 7.39399 & 3.95453 & 3.86111 & 3.86871 \\ \hline
$-0.1$ & 8.91900  & 8.91897  & 8.91897  & 6.19423 & 6.19413 & 6.19413 & 2.46999 & 2.46826 & 2.46819 \\ \hline
$0$    & 8.71735  & 8.71735  & 8.71735  & 6.00000 & 6.00000 & 6.00000 & 2.32088 & 2.32088 & 2.32088 \\ \hline
$0.1$  & 8.56817  & 8.56816  & 8.56816  & 5.87085 & 5.87079 & 5.87078 & 2.24699 & 2.24517 & 2.24532 \\ \hline
$0.5$  & 8.58903  & 8.57548  & 8.57444  & 6.09673 & 6.05237 & 6.04774 & 2.93984 & --      & 2.22017 \\ \hline
$0.9$  & 9.42112  & 9.34578  & 9.33874  & 7.16754 & 7.00056 & 6.97847 & 4.58606 & --      & 2.21706 \\ \hline
\end{tabular}
} 
\caption{Comparison of ISCO radii under TD and MP SSC, for different spin parameters $a$ and particle spin values $\sigma$, and $\hat{k}=6$. The dashes indicate that ISCOs do not exist or could not be found for those configurations.}
\label{tab:ISCO_combined_k6}
\end{table*}

\end{document}